\documentclass[prx,twocolumn,superscriptaddress, amssymb, amsmath]{revtex4-2}
\usepackage{graphicx}
\usepackage{dcolumn}
\usepackage{bm}
\usepackage{wasysym}
\usepackage{color}
\usepackage{xcolor}
\usepackage{hyperref}
\usepackage[normalem]{ulem}
\usepackage{physics}
\usepackage{relsize}
\usepackage{soul}

\newcommand{\beq}{\begin{equation}}
\newcommand{\eeq}{\end{equation}}
\newcommand{\beqn}{\begin{eqnarray}}
\newcommand{\eeqn}{\end{eqnarray}}

\newcommand{\cE}{ {\cal E} }

\newcommand{\vect}[1]{{\bm{#1}}}

\newcommand{\hrho}{\hat{\rho}}

\newcommand{\calO}{\mathcal{O}}

\renewcommand{\tr}{\mathrm{tr}}

\begin{document}

\title{Spontaneous Strong Symmetry Breaking in Open Systems: Purification Perspective}



\author{Pablo Sala}

\affiliation{Department of Physics and Institute for Quantum Information and Matter, California Institute of Technology, Pasadena, California 91125, USA}
\affiliation{Walter Burke Institute for Theoretical Physics, California Institute of Technology, Pasadena, CA 91125, USA}

\author{Sarang Gopalakrishnan}

\affiliation{Department of Electrical and Computer Engineering,
Princeton University, Princeton, NJ 08544, USA}

\author{Masaki Oshikawa}

\affiliation{Institute for Solid State Physics, University of Tokyo, Kashiwa, Chiba 277-8581, Japan }

\author{Yizhi You}

\affiliation{Department of Physics, Northeastern University, Boston, MA, 02115, USA}


\begin{abstract}
We explore the landscape of the decoherence effect in mixed-state ensembles from a purification perspective. We analyze the spontaneous strong-to-weak symmetry breaking (SSSB) in mixed states triggered by local quantum channels by mapping this decoherence process to unitary operations in the purified state within an extended Hilbert space.
Our key finding is that mixed-state long-range order and SSSB can be mapped into symmetry-protected topological (SPT) order in the purified state. Notably, the measurement-induced long-range order in the purified SPT state mirrors the long-range order in the mixed state due to SSSB, characterized by the Rényi-2 correlator. We establish a correspondence between fidelity correlators in the mixed state, which serve as a measure of SSSB, and strange correlators in the purification, which signify the SPT order. 
This purification perspective is further extended to explore intrinsic mixed-state topological order and decoherent symmetry-protected topological phases.
\end{abstract}

\maketitle
\tableofcontents
\section{Introduction}

Classifying the ground states of gapped Hamiltonians is one of the landmark achievements of many-body physics\cite{wenbook,Hastings2004,Oshikawa2000}. A central idea behind this classification is that two states are in the same phase if they are related by a finite-depth local unitary (FDLU) quantum circuit---or equivalently by a finite-time evolution under a local Hamiltonian\cite{XieChenScience,pollmann10,chen11a,chen11b,Hastings2004,hastings2011topological}. Restricting the class of allowed evolutions, e.g., by imposing symmetries, generalizes this concept to encompass ``symmetry-protected'' topological phases\cite{wangc-science,Senthil_2015,Senthil2013,ChongWang2013,ChongWang2014,Lukasz2013,XieChen2014}. The Lieb-Robinson theorem guarantees that two states that are related by an FDLU circuit have the same asymptotic correlations and entanglement structure at a large distance, and the same power to encode quantum information\cite{kitaev2003fault,nayak2008non,bravyi2010topological,bravyi2011short,hastings2011topological,lu2020detecting,yoshida2011feasibility,wildeboer2022symmetry}. The properties of a quantum state that make it valuable for quantum information processing tasks---e.g., the presence of non-abelian anyons---are robust properties that are present throughout a phase of matter, and do not depend on fine-tuning the Hamiltonian\cite{michalakis2013stability}.

The control and manipulation of these highly entangled states enable the preparation of resource states for Measurement-Based Quantum Computing (MBQC)\cite{raussendorf2007topological,eckstein2024robust,zhu2023nishimori}, where measurements on bulk qubits of a resource state facilitate universal quantum computation. In recent years, there has been a surge of progress in simulating quantum states of matter with nontrivial entanglement on platforms summarized as the noisy intermediate-scale quantum (NISQ) technology~\cite{bharti2022noisy}, including simulating exotic quantum many-body states such as topological order, spin liquids, conformal field theory quantum critical points~\cite{fang2024probing} and symmetry-protected topological (SPT) states~\cite{satzinger2021realizing,semeghini2021probing,chen2023realizing,iqbal2023creation,iqbal2023topological,tantivasadakarn2023hierarchy,baumer2023efficient,foss2023experimental}. 
A quantum state interacting with an environment can be understood as being continuously \textit{measured} by it, eventually becoming entangled with the environment's degrees of freedom. If the environmental qubits are inaccessible or the measurement outcomes are lost, this effectively leads to the tracing out of the environment's qubits, transforming the original quantum state into a mixed-state ensemble.
It is natural to ask if mixed states can be classified into distinct phases, separated (e.g.) by their ability to protect quantum information\cite{chen2023separability,chen2023symmetry,lavasani2024stability,lu2023mixed}. The threshold theorem for quantum error correction guarantees the existence of nontrivial mixed-state phases (such as the toric code subject to weak enough noise), which preserve quantum coherence\cite{Dennis_2002,lee2022decoding,LeeYouXu2022,zhu2023nishimori}. However, the tools to characterize these phases\cite{coser2019classification,bao2023mixed,fan2023diagnostics,chen2023symmetry,chen2023separability,sang2023mixed,li2024replica,zhang2022strange,lee2022symmetry,albert2014symmetries,brown2016quantum,peres1996separability,lessa2024mixed}, as well as the transitions between them, remain largely undeveloped, despite some very recent progress\cite{lee2022symmetry,ma2023topological,sohal2024noisy,dai2023steady,wang2023intrinsic,de2022symmetry,chen2024unconventional,ma2023average,hauser2023continuous,Antinucci2023,lee2024exact,lu2024disentangling}. 

The example of the noisy toric code\cite{Dennis_2002,bao2023mixed,fan2024diagnostics} illustrates a fundamental difference between mixed-state and pure-state phases. Suppose we start with one of the logical ground states of the pristine toric code, and then subject it to local noise at some rate $\gamma$, without performing any active error correction. At sufficiently short times, the mixed state generated this way will remain correctable, but at some \emph{finite} time $t^*(\gamma)$, the density of errors will exceed the capacity of even the optimal decoder to correct~\cite{Dennis_2002}: thus, logical information will become irretrievable at some finite time. Such finite-time transitions have no obvious pure-state analog. However, they seem quite generic in the context of systems subject to noise and/or measurements: a conceptually related finite-time teleportation transition has been predicted in random quantum circuits in two or more dimensions. These observations challenge our conventional understanding of what a phase of matter is in open systems~\cite{myerson2023decoherence,su2023conformal,su2023higher,lee2022symmetry,ZhangQiBi2022,ma2023topological,sohal2024noisy,chen2024unconventional,ma2023average,Antinucci2023,wang2023intrinsic,lavasani2024stability,guo2023two,wang2024anomaly}, and what properties are universal: in the pure-state context it is assumed that finite-time evolution cannot change universal properties of a phase, but whether a mixed-state is error-correctable seems like a fundamental property. Even at a technical level, the Lieb-Robinson bounds that ensure equivalence of two pure states connected by a short-depth circuit also apply to mixed states, and ensure that ``conventional'' diagnostics of order, such as correlation functions, cannot diverge at mixed state phase transitions. 
Thus, recently proposed diagnostics of mixed-state order include observables that are nonlinear in the density matrix\cite{LeeYouXu2022,fan2024diagnostics,ZhangQiBi2022,zhu2023nishimori}, metrics based on whether the density matrix can be written as an ensemble of trivial-phase pure states\cite{chen2023realizing,chen2023symmetry}, and quantities such as the conditional mutual information~\cite{sang2024stability, sang2023ultrafast, zhang2024nonlocal, lee2024universal} (and the related notion of measurement-induced entanglement~\cite{lu2023mixed, cheng2023universal}). However, the physical interpretation of these diagnostics, and their relation to concepts of pure-state order, remains opaque.

The noisy toric code is an example where an ordered phase appears to \emph{lose} long-range correlation under a finite-depth local quantum channel. The reverse phenomenon can also occur, and has been termed ``spontaneous strong symmetry breaking'' (SSSB)~\cite{LeeYouXu2022,ma2023topological} where the decoherence effect triggers long-range ordering in mixed states. SSSB occurs in product states\footnote{More generally, weakly entangled states that are all themselves reachable from product states by finite-depth local channels (of which finite-depth local unitaries are a subclass).} subject to finite-depth local channels, so from the point of view of conventional correlation functions a mixed state possessing SSSB will appear trivial. Meanwhile, observables that are nonlinear in the density matrix, such as the Renyi-2 correlator and quantum relative entropy, behave singularly~\cite{LeeYouXu2022, ma2023average, fan2023diagnostics}. 
Intuitively, an open system is said to possess a strong symmetry when the system-environment interaction does not exchange symmetry charges---e.g., an open system of electrons coupled to phonons has a strong fermion parity symmetry. A pertinent question is why spontaneous strong symmetry breaking can be triggered by decoherence and how to characterize such unique long-range order in an open quantum system.



The goal of this work is to characterize spontaneous strong-to-weak symmetry breaking (SSSB) from the perspective of purification, specifically by considering the mixed-state density matrix as a partial trace of a pure state in an extended Hilbert space. From this viewpoint, quantum channels acting on the mixed-state ensemble are equivalent to unitary operators acting on the purified state. Our central finding is that mixed-state long-range order and spontaneous strong-to-weak symmetry breaking, induced by finite-depth quantum channels, can be mapped to symmetry-protected topological (SPT) order in the purified framework. This approach highlights why mixed-state long-range order can manifest within a finite time: an SPT phase transition from a trivial phase is achievable through a symmetry-breaking finite-depth circuit. Particularly, we demonstrate that SSSB in a mixed state implies that its purification exhibits a non-vanishing strange correlator, a hallmark of an SPT wavefunction. We link these strange correlators to previously discussed observables, including the fidelity correlator and type-2 strange correlator. Furthermore, we establish a correspondence between the mixed-state Rényi-2 correlator and measurement-induced order in the corresponding purified state.

\begin{figure}[h!]
\includegraphics[width=0.5\textwidth]{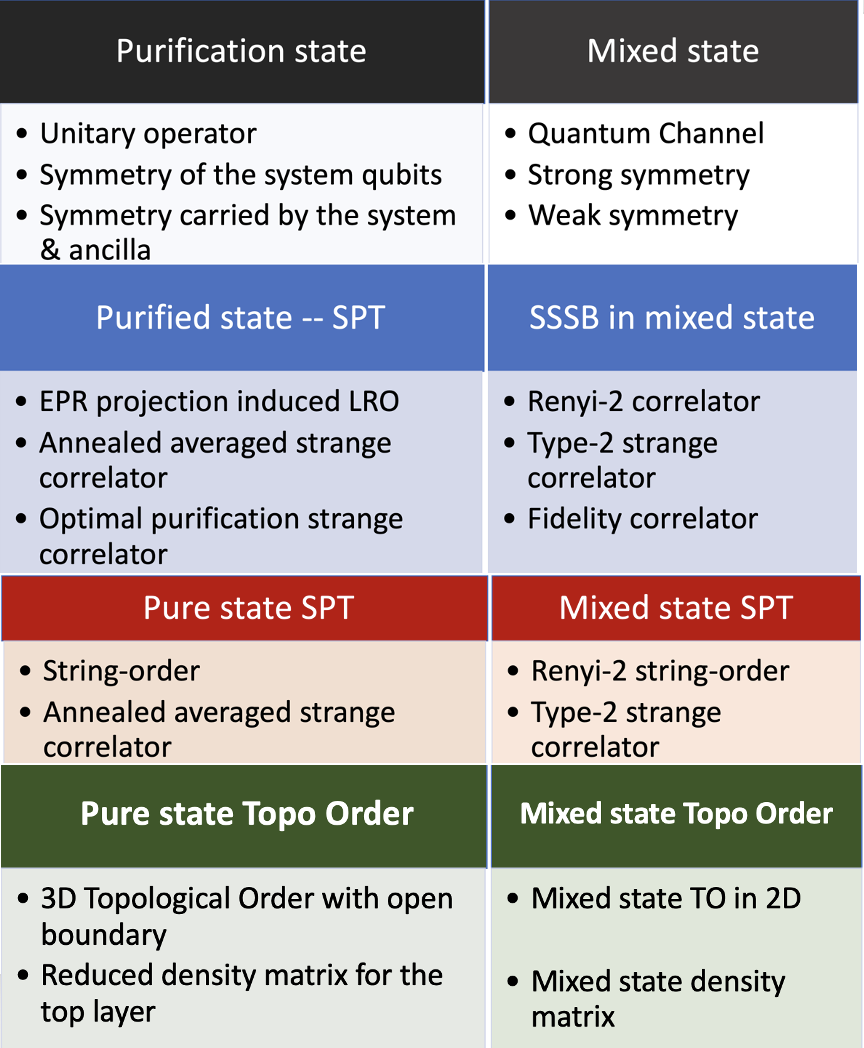}
\caption{\textbf{Summary of results.} Mapping between exotic mixed state phenomena and its purification description. }
\label{all}
\end{figure}

This work is organized as follows: In Sec.~\ref{sec:purifiedspt}, we introduce strong-to-weak \(\mathbb{Z}_2\) symmetry breaking in mixed states triggered by local quantum channels in both 1D and 2D. These states can be mapped to a SPT phase within the purification framework. We trace the Rényi-2 correlator of SSSB back to the measurement-induced long-range order observed in the purified SPT state. Additionally, we find that while SSSB is unstable in 1D when subject to a finite measurement rate quantum channel, it becomes significantly more stable up to a certain threshold in 2D and higher dimensions. In Sec.~\ref{sec:strange}, the correspondence between the averaged strange correlator in the purified state and various SSSB observables in the mixed state is demonstrated. In Sec.~\ref{sec:generalsym}, we broaden the scope of SSSB to include a wider array of symmetry groups, such as higher-form symmetry, subsystem symmetry, and continuous groups. Finally, in Secs.~\ref{sec:topo} to \ref{sec:SPT}, we explore mixed-state topological orders and SPT phases from a purification perspective. Fig.~\ref{all} summarizes the main results of this work in a comprehensive table.

\section{General formalism}\label{general}

\subsection{Mixed states, quantum channels and purifications}

We begin with a many-body pure state \( |\psi \rangle \), defined on the system Hilbert space $\mathcal{H}$. Under open-system dynamics, the system is then coupled to an environment Hilbert space $\mathcal{A}$, which (without loss of generality) we can take to also be prepared in a fixed pure state $\ket{0}$. We will follow the standard practice of referring to these environment qubits (or more generally qudits) as ancillae. The global evolution is a unitary map $ \mathcal U$ defined on $\mathcal{H} \otimes \mathcal{A}$, such that $\mathcal U (\ket{\psi} \otimes \ket{0}) = \ket{\Psi_p} \in \mathcal{H} \otimes \mathcal{A}$. In general, some or all of the ancillae are inaccessible after this process. To describe the accessible degrees of freedom in the system, one traces out $\mathcal{A}$, arriving at the density matrix $\hrho^D \equiv \mathrm{Tr}_{\mathcal{A}} (\ket{\Psi_p} \bra{\Psi_p})$. We mainly focus on the cases in which the initial state $\ket{\psi}$ is a product state, the system and ancilla are defined on a geometrically local space, and the operator $U$ can be written as a finite-depth, geometrically local unitary circuit. 

The process we described above is one way of defining a quantum channel, i.e., a map $\mathcal{E}$ between legal density matrices on $\mathcal{H}$. The perspective where the channel is treated as a unitary in a larger Hilbert space is called the Stinespring form of the channel. The pure state $\ket{\Psi_p} \in \mathcal{H} \otimes \mathcal{A}$ is called a purification of $\hrho^D$. Stinespring forms and purifications are not unique, as $\hrho^D$ is invariant if one replaces the map $\mathcal U$ above with a map $(\mathbb{I}_{\mathcal H} \otimes W_{\mathcal A}) \mathcal U$, where $W$ is an arbitrary unitary that acts only on the environment $\mathcal{A}$~\cite{preskill1998lecture}. 

%
%
Maps between legal density matrices are known as completely positive trace-preserving (CPTP) maps, and take the general form $\hrho^D = \mathcal{E}(\hrho) = \sum_m K_m \hrho_0 K_m^\dagger$, where \(\{K_m\}\) is a set of ``Kraus operators'' satisfying the trace-preserving condition \(\sum_m K_m^\dagger K_m = \mathbb{I}_{\mathcal{H}}\). Here the number of Kraus operators match the local dimension of the ancillary space.
As we saw before, to derive the Kraus form from the Stinespring form, one traces out over the ancillary space $\mathcal{A}$, such that $K_m=\langle 0|U^\dagger|m\rangle$ with $\{\ket{m}\}$ an orthonormal basis in $\mathcal{A}$.
When the Stinespring form of a channel involves a local unitary $\mathcal U$ as specified above, all the Kraus operators can be chosen to be local operators. 

\subsection{Weak and strong symmetries}

Recall that for a pure state, a symmetric wave function implies that the quantum state carries a conserved charge with respect to the symmetry \(G\), thereby rendering it invariant under the symmetry transformation \(U_g\)
\begin{align}
    U_g |\Psi \rangle= e^{i\theta}|\Psi \rangle,
\end{align}
with $e^{i\theta}$ being a global phase for all $g\in G$ 

For a mixed state, there are two notions of symmetry---``weak'' and ``strong''. The concept of weak and strong symmetries was first introduced for channels (or equivalently, their continuous-time version, Lindblad master equations)~\cite{Bu_a_2012,PhysRevA.89.022118}. However, as we will discuss below, these can also be seen as induced properties of mixed-state density matrices. We first explain them in the purification (Stinespring) picture. In this picture, a weak symmetry exists when the system-environment unitary $\mathcal U$, and the initial state of the environment $\ket{0}$, are invariant under a symmetry transformation that acts on $\mathcal{H} \otimes \mathcal{A}$. Thus, for example, if a system exchanges particles with the ancilla but the total number of particles in the purified state defined on $\mathcal{H} \otimes \mathcal{A}$ is conserved, that corresponds to a weak $U(1)$ symmetry. On the other hand, a strong symmetry requires that the symmetry operator acts only on \(\mathcal{H}\). That is, when the system interacts with the environment, there is no charge exchange between the system and the ancilla.
When a channel \( \mathcal{E} \) is invariant under a strong symmetry, it means that each Kraus operator commutes with the symmetry operation, so \( [K_m, U_g] = 0 \) for all \( g \in G \).

Instead of regarding these symmetries as properties of quantum channels, we can instead treat them as properties of a \emph{mixed state} $\hrho$, in particular one that was arrived at by applying $\mathcal{E}$ to a product state invariant under $G$~\cite{LeeYouXu2022,ma2023topological,ma2023average,zhang2022strange}. The constraints on $\hrho$ are inherited from those on $\mathcal{E}$. A weak symmetry requires that the density matrix remains invariant under the symmetry transformation \(U_g\), acting on both the left (ket) and right (bra) parts of the density matrix.
\begin{align}
    \hrho= U_g \hrho U^\dagger_g.
\end{align}
The weak symmetry transformation can be interpreted as implementing the symmetry operation on both the ket and bra spaces of the density matrix. Physically, it requires that the density matrix be block-diagonal, with each block corresponding to a different charge under \(G\). This is intuitive since the system can exchange charge with the bath; however, as the charge is conserved for the system and ancilla as a whole, the reduced density matrix of the system can still be block-diagonal in different charge sectors. For the special case where the density matrix is the partition function in thermal equilibrium \( \hrho = e^{-\beta H} \), the invariance of \( \hrho \) under the weak symmetry \( G \) implies that the Hamiltonian is \( G \)-symmetric.

Density matrices with strong symmetry, on the other hand, are those with a \emph{definite} value of the charge. When a quantum channel has a strong symmetry, it preserves the invariance of density matrices under the \( U_g \) symmetry transformation which acts exclusively on either the left or the right part of the density matrix:
\begin{align}
     e^{i\theta}\hrho= U_g \hrho 
\end{align}
with $e^{i\theta}$ being a global phase for all $g\in G$.
If we diagonalize the density matrix as \( \hrho = \sum_{\Lambda} p_{\Lambda} |\Lambda \rangle \langle \Lambda| \), the strong symmetry condition requires that all eigenvectors of the density matrix also be eigenstates of the symmetry \( G \), such that \( U_g |\Lambda \rangle = e^{i\theta} |\Lambda \rangle \), each carrying the same symmetry charge.

\subsection{Strong-to-Weak Symmetry Breaking (SSSB) driven by finite-depth quantum channels}

We begin by reviewing a concrete example of spontaneous $\mathbb{Z}_2$ SSSB, driven by local quantum channels, initially introduced in Ref.~\cite{LeeYouXu2022,lee2022symmetry,ma2024symmetry}.

Before proceeding, we briefly revisit the correlation functions of symmetry breaking in thermal equilibrium ensembles. For canonical ensembles \( \hrho = e^{-\beta H} \) of systems with conserved charge $G$, both the Hamiltonian and the thermal density matrix exhibit invariance under the weak symmetry,
\begin{align}
    H=U_g H U^{\dagger}_g, ~\hrho=U_g\hrho U^{\dagger}_g
\end{align}
With the symmetry operator acting on both the left and right parts of the density matrix, the spontaneous breaking of weak symmetries~\footnote{Following Ref.~\cite{ma2023topological}, we characterize spontaneous symmetry breaking on a finite system by the presence of long-range order as measured by symmetric correlations.} is evidenced by the non-vanishing two-point function \(\Tr[\hrho O(x)O^{\dagger}(y)]\) for a given charged operator \(O(x)\).

Now we move on to the strong symmetry condition. For any mixed-state density matrix invariant under the strong symmetry \(U_g\), the charge operator \(O(x)\) acting on the doubled density matrix should vanish\footnote{Assume that $U_g O(x) U_g^\dagger = e^{iq} O(x)$ and that $U_g \hrho = e^{i\alpha}\hrho$. Then $\Tr\left( O(x)\hrho O^\dagger(x) \hrho\right) =e^{iq}\Tr\left( O(x)\hrho O^\dagger(x) \hrho\right) $, which for $q\neq 2\pi $ implies $\Tr\left( O(x)\hrho O^\dagger(x) \hrho\right)=0$. }:
\begin{align}
    \frac{\Tr(O(x)\hrho O^{\dagger}(x)\hrho)}{\Tr(\hrho^2)}=0
\end{align} 
Notably, the operators \(O(x)\) and \(O^{\dagger}(x)\) act simultaneously on the ket and bra spaces of the density matrix. Such operations are `charged' under strong symmetry but remain neutral under weak symmetry. Prompted by this observation, to define spontaneous SSSB, we aim to identify the correlation function of an operator that is charged under strong symmetry but neutral under weak symmetry. 

The non-vanishing correlation function indicative of SSSB manifests as four-point functions acting on the doubled density matrix:
\begin{equation}
\label{eq:swSSB}
    \frac{{\rm tr}\left(O(x)O^\dag(y)\hrho O(y) O^{\dagger}(x)\hrho \right)}{{\rm tr}(\hrho^2)} =C^{\text{II}}(x,y)
\end{equation}
For some charged operators \(O(x)\) and \(O(y)\) with \(|x-y| \to \infty\), Eq.~\ref{eq:swSSB}, also referred to as the Rényi-2 correlator in Ref.~\cite{LeeYouXu2022,ma2023topological} or the Edwards-Anderson correlator, is widely used in systems with localization to characterize symmetry breaking within disordered ensembles. As an immediate sanity check, we observe that such strong-to-weak SSB is distinctive in mixed states since, for pure states, the Rényi-2 correlator is merely the square of the conventional correlator.

The Rényi-2 correlator can be interpreted as the correlation function for the paired operators \( O(x) \) and \( O^{\dagger}(x) \), which act simultaneously on the ket and bra spaces, thereby creating a strong symmetry charge at site \( x \). Ref.~\cite{LeeYouXu2022,ma2023topological} presents a straightforward illustration of SSSB by analyzing a mixed state of the 1D spin chain, denoted as \( \hrho_+ = I + \prod_i X_i \), which demonstrates strong \( \mathbb{Z}_2 \) symmetry generated by \( X \equiv \prod_i X_i \), where \( X_i \) denotes the Pauli-x operator at each site \( i \). To reach this mixed state through local quantum channels, consider an initial state with all qubits polarized in the \( S_x \) direction, expressed as \( |\phi_0\rangle = \otimes_i |\rightarrow\rangle_i \). To initiate the strong-to-weak symmetry breaking, one implements a quantum channel that ``measures'' the spin bilinear term \( Z_i Z_{i+1} \) at every bond \( (i,i+1) \) as:
\begin{align}
    & \hrho_+ = \mathcal{E}[\hrho_0], \ \ \mathcal{E} = \prod_{\vect{i}} \cE_{\vect{i}}, \nonumber \\ &  \cE_{\vect{i}}[\hrho_0] =
    \frac{1}{2} \hrho_0 + \frac{1}{2}  Z_i Z_{i+1}\hrho_0 Z_i Z_{i+1}. 
\end{align}
The resulting mixed state \(\hrho_+\) spontaneously breaks a strong \(Z_2\) symmetry to a weak \(\mathbb{Z}_2\) symmetry, evidenced by the non-vanishing correlation outlined in Eq.~\eqref{eq:swSSB}.
 \begin{equation}
    \frac{\Tr \left(Z_0 Z_i\hrho_+ Z_0 Z_i\hrho_+ \right)}{\Tr(\hrho_+^2)}  =1
    \label{doubleco}
\end{equation}
When expressed in the \(Z\)-basis, \(\hrho_+\) is essentially a convex sum of GHZ states:
\begin{equation}\label{densitybreak}
    \hrho_+ \sim \sum_{s} (|s\rangle+X|s\rangle)(\langle s|+\langle s|X),
\end{equation}
where \(s\) is a bit string in the \(Z\)-basis.

\section{Mixed state SSSB from SPT purification}\label{sec:purifiedspt}

In this section, we explore the strong to weak symmetry breaking in mixed states triggered by local quantum channels, from the perspective of purification. Just as all mixed states can be viewed as subsystems of a pure state (denoted as the \textit{purification state}) in an extended Hilbert space, local quantum channels can equivalently be described by local unitary operators acting on the purification state. While the combined ancilla and system qubits remain in a pure state after the unitary operations, the system's density matrix, \( \hrho \), obtained by tracing out the ancilla, generally represents a mixed state. 

As the initial purified state (encompassing both the system and the ancilla) lacks long-range order, its quantum correlations remain short-ranged within the Lieb-Robinson bound after applying finite-depth local unitaries. Consequently, we do not expect the system's density matrix, after tracing out the ancilla, to exhibit any long-range order measurable by physical observables linear in \(\hrho\). Meanwhile, Ref.~\cite{LeeYouXu2022,ma2023topological,ma2024symmetry}, as summarized in Eq.~\ref{densitybreak}, provides a straightforward yet illustrative example of how a local quantum channel could induce strong symmetry breaking, exemplified by the Rényi-2 correlator in Eq.~\ref{doubleco}. A pertinent question arises: If the mixed state \(\hrho\) undergoes spontaneous strong-to-weak symmetry breaking induced by quantum channels, what happens to its purification states under unitary evolution? Additionally, what characteristics must the purification state possess to allow SSSB and long-range Rényi-2 correlation for the system as a mixed state, after tracing out the ancilla?

\subsection{1D SSSB: Purification from SPT state}

To set the stage, we focus on the purified wave function within the extended Hilbert space, which encompasses both system and ancilla qubits. Although we will eventually trace out the ancilla qubits to obtain the mixed-state density matrix, this approach aims to explore the SSSB of mixed states from the perspective of purification and reveal the correspondence between conditional long-range correlation in the purified state and the long-range order manifested via the Rényi-2 correlator of the mixed state.

\begin{figure}[h!]
\includegraphics[width=0.45\textwidth]{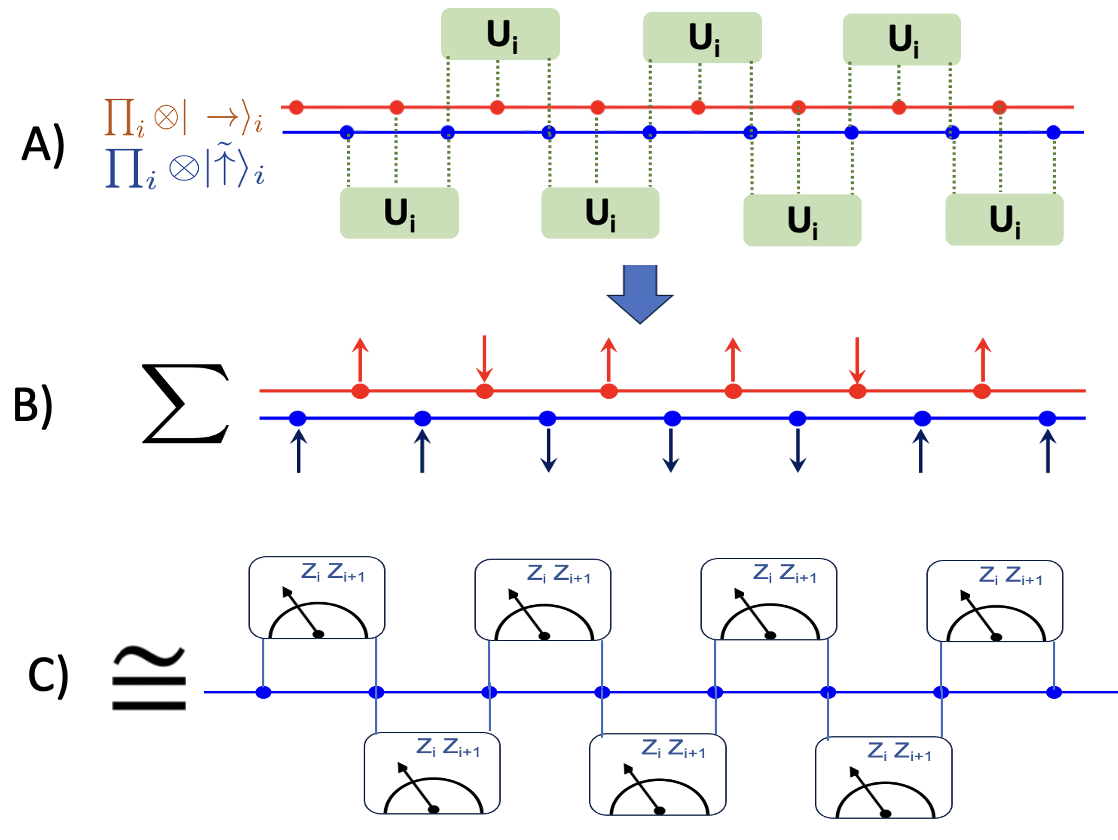}
\caption{\textbf{$1$D SSSB and purification.} A) The unitary operator entangles the system (blue) with the ancilla (red). The unitary gates consist of a cluster of three-body gates, each acting on two adjacent system qubits and one ancilla situated between them.
B) The unitaries fix the total \(S_z\) parity of the two system spins on each link and the intervening ancilla spin, ensuring \(Z_i \tilde{Z}_i Z_{i+1} = 1\).
C) Such a unitary in the extended Hilbert space is equivalent to quantum channels that measure \(Z_i Z_{i+1}\) on adjacent spins.}
\label{1d}
\end{figure}

The system of interest encompasses a 1D spin system, initialized as \( |\phi_0\rangle =  \otimes_i |\rightarrow \rangle_i \), along with a chain of ancillae that resides on the links between the system's spins, as depicted in Fig.\ref{1d}.
The initial state of the ancillae is a tensor product of spins, polarized in the \( S_z \) direction, expressed as \( |\phi^A_0\rangle =  \otimes_i |\tilde{\uparrow} \rangle_i \). The unitary operator \( U \) has the following form:
\begin{align}
\label{unitary1}
U&=\prod_i U_{i,i+1},~\\
   \nonumber U_{i,i+1}&=\frac{(1+Z_i Z_{i+1})}{2} \tilde{I}_{i,i+1}+\frac{(1-Z_iZ_{i+1})}{2} i\tilde{Y}_{i,i+1}
\end{align}
The unitary operator involves a cluster of three-body gates, each acting on two adjacent system qubits and one ancilla situated between them, as illustrated in Fig.\ref{1d}.
Here, \( Z_i \) refers to the operator acting on the system qubits, while \( \tilde{Y}_{i,i+1} \) acts on the ancilla. Such a unitary operation exhibits a \textit{decorated domain wall} feature. When the system spins at the \((i,i+1)\) link are aligned parallel in the \(z\)-direction, the \((i,i+1)\) ancilla on the link remains in the \( |\tilde{\uparrow} \rangle_{i,i+1} \) state. Conversely, if the system spins at the \((i,i+1)\) link are opposite in the \(z\)-direction, ancilla at  \((i,i+1)\) is flipped to the \( |\tilde{\downarrow} \rangle_{i,i+1} \) state. 
After applying the local unitary, the resultant wave function in the extended Hilbert space precisely reflects a symmetry-protected topological (SPT) state,

\begin{align}\label{zxz} 
|\Psi_{\text{SPT}} \rangle=U\bigotimes_{\textrm{sites }i} |\rightarrow \rangle_i \bigotimes_{\textrm{bonds } (i,i+1)} |\tilde{\uparrow} \rangle_{(i,i+1)}
\end{align}

which corresponds to the ground state of a stabilizer Hamiltonian,
\begin{align}
H=- \sum_i \left(Z_i \tilde{Z}_{i,i+1} Z_{i+1} + \tilde{X}_{i-1,i} X_i  \tilde{X}_{i,i+1} \right)
\label{1Dstabilizer}
\end{align}
This Hamiltonian comprises two stabilizer operators. The $Z_i \tilde{Z}_{i,i+1} Z_{i+1} $ stabilizer ensures that the $S^z$ parity in the 3-body clusters is even, while the $\tilde{X}_{i-1,1} X_i  \tilde{X}_{i,i+1}$ stabilizer generates resonance between distinct spin patterns that share the same $S^z$ parity across all clusters. The stabilizer Hamiltonian, along with the wave function, respects the following symmetry:
\begin{align}
\mathbb{Z}_2^A: \prod_i \tilde{Z}_{i,i+1},~~
\mathbb{Z}_2^S: \prod_i X_i.
\label{sym}
\end{align}
The \(\mathbb{Z}_2^A\) symmetry acts on the ancilla, while \(\mathbb{Z}_2^S\) acts on the system. Remarkably, the unitary operator applied in Eq.~\ref{unitary1} preserves the \(\mathbb{Z}_2^S\) symmetry while breaking the \(\mathbb{Z}_2^A\) symmetry. This outcome is expected, as no local unitary path can connect a tensor product state with an SPT state while preserving both symmetries. While the SPT wave function in Eq.~\ref{zxz} is short-range correlated, it carries hidden quantum correlations that can be characterized by a non-vanishing string order parameter:
\begin{align}
\langle O_s \rangle= \langle \Psi_{\text{SPT}} |Z_i ~(\prod^{n}_{a=0} \tilde{Z}_{i+a,i+1+a} ) ~Z_{i+n+1}| \Psi_{\text{SPT}} \rangle=1
\end{align}
This string order reveals the conditional mutual information shared between the system and the ancilla (see \cite{verresen2021efficiently}). The two-point correlation of the system spins, denoted as \(\langle Z_i Z_{i+n+1} \rangle\), is determined by the \(S^z\) parity charge of the ancilla on the string \(\prod^{n}_{a=0} (\tilde{Z}_{i+a,i+1+a})\) in between. Therefore, if we project all ancillae into the \(|\tilde{\uparrow}\rangle\) state, the system would transition into a spontaneously symmetry-breaking `cat-state' with long-range order (LRO) for the two-point correlation \(\langle Z_i Z_{i+n+1} \rangle=1\).

Once we trace out the ancilla and obtain the system's density matrix \(\hrho = \Tr_{\text{ancilla}}|\Psi_{\text{SPT}} \rangle \langle \Psi_{\text{SPT}}|\), it indeed reproduces the mixed state introduced in Eq.~\ref{densitybreak}. This process averages out the even versus odd \(S^z\) charge of the ancilla on the string. Consequently, the resultant two-point correlation vanishes, as indicated by \(\Tr[\hrho Z_i Z_{i+n+1}] = \langle \Psi_{SPT}|Z_i Z_{i+n+1}|\Psi_{SPT}\rangle =0\). However, if we examine the Rényi-2 correlator introduced in Eq.~\ref{doubleco}, based on two copies of the density matrix, it remains non-vanishing at large distances indicating SSSB.

\begin{figure}[h!]
\includegraphics[width=0.45\textwidth]{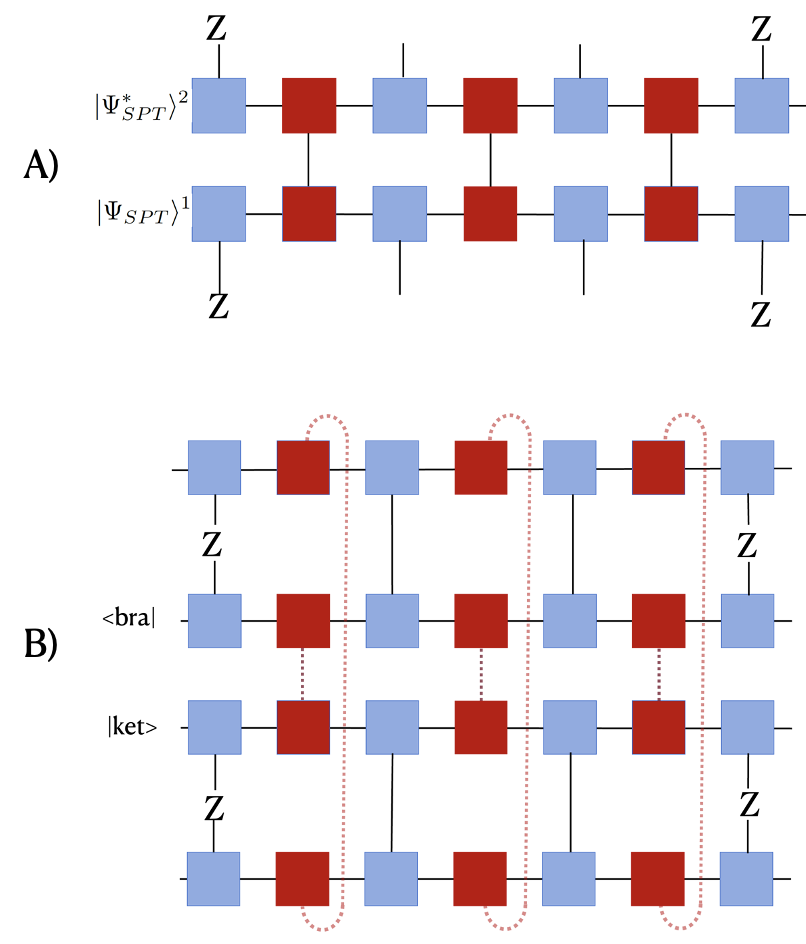}
\caption{\textbf{R{\'e}nyi-II correlator and EPR-induced long range order.} A) Taking two copies of the SPT state \(|\Psi_{\text{SPT}}\rangle^{1} \otimes |\Psi_{\text{SPT}}\rangle^{2}\) and projecting the ancilla (indicated by the red square) from both copies onto a symmetric EPR pair, the post-projection state of the system (indicated by the blue square) displays a long-range correlation characterized by the four-point correlator.
B) The tensor representation of the Rényi-2 correlator measures the four-point correlator acting on the double-density matrix. When we trace out the legs of the ancilla (tensor legs contraction), it effectively projects the ancilla in the bra and ket space onto the EPR pair.}
\label{wf}
\end{figure}

To provide a physical interpretation of the Rényi-2 correlator, we duplicate our entire Hilbert space by creating two identical copies of the SPT state, denoted \(|\Psi_{\text{SPT}}\rangle^{1}\) and \(|\Psi_{\text{SPT}}\rangle^{2}\).
Next, we perform a Schmidt decomposition between the ancilla and the system for each copy of the SPT wave function as:
\begin{align}
|\Psi_{SPT}\rangle^{1}= \sum_{\alpha} \lambda_{\alpha}|\alpha \rangle^1_s|\tilde \alpha \rangle^1_a, ~|\Psi_{SPT}\rangle^{2}= \sum_{\alpha} \lambda^*_{\alpha}|\alpha^* \rangle^2_s|\tilde \alpha^* \rangle^2_a
\end{align}
Here, we take the complex conjugate of the second copy of the wave function, denoted by \(|\Psi_{\text{SPT}}\rangle^{2}\). Its physical interpretation and essentiality will be explained shortly. \(|\alpha \rangle_s\) denotes the Schmidt basis for the system, while \(|\tilde{\alpha} \rangle_a\) refers to the Schmidt basis for the ancillae. \(|\alpha^* \rangle\) refers to the complex conjugate of the vector in some fixed basis.
These two identical copies, \(|\Psi_{\text{SPT}}\rangle^{1} \otimes |\Psi_{\text{SPT}}\rangle^{2}\), carry a non-vanishing string order, which is essentially the product of the strings for each copy:
\begin{align}
&\langle O^1_s O^2_s \rangle
\\ \nonumber &= 
\langle Z^1_i Z^2_i ~(\prod^{n}_{a=0} \tilde{Z^1}_{i+a,i+1+a} \tilde{Z^2}_{i+a,i+1+a})~ Z^1_{i+n+1}Z^2_{i+n+1} \rangle
 \label{stringdual}
\end{align}
We now project ancilla from the first and second copy onto a symmetric EPR pair, hence forcing their alignment in the \(S^z\) direction:
\begin{align}
    \hat{P}_{i,i+1} = \frac{1}{2}\left(|\tilde{\uparrow}^1 \tilde{\uparrow}^2\rangle +|\tilde{\downarrow}^1 \tilde{\downarrow}^2 \rangle\right)
    \left(\langle \tilde{\uparrow}^1 \tilde{\uparrow}^2| +\langle\tilde{\downarrow}^1 \tilde{\downarrow}^2|\right)_{(i.i+1)}
\end{align}
This projection is implemented for every ancilla pair at bond $(i,i+1)$. The normalized wavefunction, after this projection, is given by:
\begin{align}\label{postp}
&|\Psi \rangle_{pp} \sim \prod_i \hat{P}_{i,i+1} [(\sum_{\alpha} \lambda_{\alpha}|\alpha \rangle^1_s|\tilde \alpha \rangle^1_a) (\sum_{\alpha'} \lambda^*_{\alpha'}|{\alpha^*}' \rangle^2_s|\tilde {\alpha^*}' \rangle^2_a)]\nonumber\\
&\rightarrow |\Psi \rangle_{pp}=\frac{1}{\sqrt{\sum_{\alpha} |\lambda_{\alpha}|^4}}\sum_{\alpha} |\lambda_{\alpha}|^2~|\alpha \rangle^1_s |\alpha^* \rangle^2_s 
\end{align}
In the last step, we omit the ancilla qubits as they form a tensor product of EPR pairs between two copies and are decoupled to the system qubits.
The projection \(\hat{P}_{i,i+1}\) on the two ancilla copies results in the charge string \((\prod^{n}_{a=0} \tilde{Z}^1_{i+a} \tilde{Z}^2_{i+a})\) being uniformly even throughout the post-projected wave function. Consequently, the post-projection state \(|\Psi \rangle_{pp}\) exhibits long-range order in the four-point correlation function:
\begin{align}\label{mio}
\langle \Psi |_{pp}
~Z^1_i Z^2_i  Z^1_{i+n+1}Z^2_{i+n+1}|\Psi \rangle_{pp}=1
\end{align}

Alternatively, we could trace out the ancilla from the purified state in Eq.~\eqref{zxz} obtaining the density matrix $\hrho= \sum_{\alpha} |\lambda_{\alpha}|^2|\alpha \rangle \langle \alpha |$.
Tracing out the ancilla effectively involves projecting the ancilla in both the ket and bra spaces to be identical.
By considering the bra vector as a duplicate copy, the ancilla tracing procedure exactly corresponds to the projection operation in Eq.~\ref{postp}. 
Drawing from this analogy, the Rényi-\(2\) correlator precisely corresponds to the four-point correlation of the post-projected wave function (see Appendix.~\ref{app:pp} for detailed derivation):
 \begin{align}
    &\frac{\Tr \left(Z_0 Z_i\hrho Z_0 Z_i\hrho \right)}{\Tr (\hrho^2)} =\langle \Psi |_{pp}
Z^1_i Z^2_i  Z^1_{i+n+1}Z^2_{i+n+1}|\Psi \rangle_{pp}
\end{align}
This result demonstrates that the Rényi-\(2\) correlator can be interpreted as the correlation function of the post-projection state \( |\Psi \rangle_{pp} \). If we conceptualize the density matrix from the tensor perspective as illustrated in Fig.\ref{wf}, tracing over the density matrix entails connecting the ancilla's legs between the ket and bra spaces. This process is analogous to taking two copies of the SPT state \( |\Psi_{\text{SPT}}\rangle^{1} \otimes |\Psi_{\text{SPT}}\rangle^{2} \) and projecting the ancilla from both copies onto a symmetric EPR pair. In this context, we consider the second copy of the wave function in its complex conjugate form to mirror the wave function in the bra space.
This correspondence implies that measuring any operators on the duplicated density matrix is equivalent to duplicating the purified state as \( |\Psi_{\text{SPT}}\rangle^{1} \otimes |\Psi_{\text{SPT}}\rangle^{2} \), followed by an EPR projective measurement to ensure the ancilla in both copies are identical. While local unitary operators acting on the system and ancilla (with its duplicate) cannot induce long-range order (LRO), an additional projective measurement can facilitate this magic. This phenomenon of measurement-induced long-range order was originally proposed in Ref.~\cite{Briegel_01, Raussendorf_05,Bolt_16,piroli2021quantum,verresen2021efficiently,tantivasadakarn2021long,lu2022measurement,lu2023mixed,bravyi2022adaptive} as a shortcut to create the long-range entangled state.
The emergence of long-range order in the Rényi-\(2\) correlator is attributed to the non-vanishing measurement-induced long-range order in its purified state. By projecting the ancilla charges on the string into the even sector, the post-projection state manifests long-range order.

\subsection{General quantum channel in 1D}

We now delve into a broader scenario of quantum channels with a tunable error rate \(p\). These can be interpreted as the system being measured at a given rate, but the outcomes of the measurement not being recorded, effectively resulting in dephasing noise. This type of decoherence channel can be represented as follows:
\begin{align}\label{decohere_1general}
    & \hrho^D = \mathcal{E}[\hrho_0], \ \ \mathcal{E} = \prod_{i,i+1} \cE_{i,i+1}, \nonumber \\ &  \cE_{\vect{i}}[\hrho_0] = (1 - p) \hrho_0 + p Z_i Z_{i+1} \hrho_0 Z_i Z_{i+1}. 
\end{align}
The quantum channel \(\mathcal{E}\) preserves the strong \(\mathbb{Z}_2\) symmetry. When \(p = 1/2\), the decoherence channel becomes a pure measurement channel, leading to the mixed state $\rho_{+}$ introduced in Eq.~\ref{densitybreak} that breaks the strong \(\mathbb{Z}_2\) symmetry spontaneously. 
A pertinent question arises: can spontaneous strong-to-weak symmetry breaking occur at a finite \(p\)? More specifically, is it possible to continuously change the error rate and trigger a phase transition?

Our previous discussion highlighted that the long-range order in the mixed state's Rényi-2 correlator is inherited from the measurement-induced long-range order of the purified SPT state in the enlarged Hilbert space. Therefore, it would be beneficial to examine the corresponding unitaries acting on the enlarged Hilbert space, defined as follows:
\begin{align}\label{unitary2}
U(\theta)=&\prod_i U_{i,i+1}(\theta),~\nonumber\\
    U_{i,i+1}(\theta)=&\frac{(1+Z_i Z_{i+1})}{2} (\cos(\theta) \tilde{I}_{i,i+1}+i\sin(\theta) \tilde{Y}_{i,i+1})~\nonumber\\
    &+\frac{(1-Z_iZ_{i+1})}{2} (\sin(\theta) \tilde{I}_{i,i+1}+i\cos(\theta)\tilde{Y}_{i,i+1}) \nonumber\\
    |\Psi_{\text{SPT}}(\theta) \rangle&=U(\theta)~ \prod_i \otimes |\rightarrow \rangle_i \otimes |\tilde{\uparrow} \rangle_{i,i+1}
\end{align}
When \(\theta=0\), the gate simplifies to the unitary introduced in Eq.~\ref{unitary1}, which corresponds to the measurement-only channel. At \(\theta=\pi/4\), the unitary only rotates the ancilla qubits, leaving the system qubits untouched, ensuring that the system remains in a pure state after tracing out the ancilla. Upon applying \(U(\theta)\), the purified state $\Psi_{\text{SPT}}(\theta)$ in the extended Hilbert space retains \(\mathbb{Z}_2^A\) symmetry (act on the ancillae) in Eq.~\ref{sym} only when \(\theta=0\). 
Returning to the quantum channel perspective, after applying this unitary and tracing out the ancilla, we return to the general quantum channel in Eq.~\ref{decohere_1general}, with an error rate $p=\frac{1-\sin(2\theta)}{2}$. 

For \(\theta\) within the interval \((0, \pi/4]\), the wave function \(\Psi_{\text{SPT}}(\theta)\) breaks \(\mathbb{Z}_2^A\) symmetry, leading to the immediate disappearance of the measurement-induced long-range order\cite{tantivasadakarn2021long,verresen2021efficiently}. As a result, the EPR-projected wave function defined as Eq.~\ref{mio} decays exponentially for finite values of \(\theta\), causing the Rényi-2 correlator to exhibit only short-range correlations. In appendix.~\ref{app:ising}, we demonstrate that the Rényi-2 correlator for \(p < 1/2\) maps to the correlation function of the 1D Ising model at finite temperature, which lacks long-range order. Consequently, SSSB in 1D is fragile and can only occur in pure measurement channels at \(p = 1/2\). In our next section, we will show that SSSB can be more robust in a 2D quantum channel, with a phase transition occurring at a finite measurement rate.

The correspondence between local quantum channels in the mixed state and local unitary operations in the purified state provides new insights into our exploration of spontaneous strong-to-weak symmetry breaking:
(i) The presence of mixed-state SSSB, detectable via the Rényi-2 correlator, stems from the conditional mutual information shared between the system and the ancillae in the SPT state \cite{cheng2023universal}. This raises a compelling question: Can all mixed-state long-range orders be purified as an SPT wave function?
(ii) If a quantum channel induces an SSSB transition in a mixed state, what occurs in its purified state during the transition?
In what follows, we will adopt a comprehensive approach and demonstrate how various examples of SSSB can be mapped to SPT through purifications.

\subsection{2D SSSB: Purification from 1-form SPT state}\label{sec:2d}

We now proceed to investigate a concrete example of SSSB on a 2D square lattice. As before, we begin our analysis with the purified state defined in the extended Hilbert space. The purified state comprises the system qubits located at the vertices, initialized as \( |\phi_0\rangle = \otimes_i |\rightarrow \rangle_i \), together with ancilla qubits living on the links. The initial state of these ancillae \( |\phi^A_0\rangle = \otimes |\tilde{\uparrow} \rangle \), is a tensor product of spins polarized in the \( S_z \) direction. To entangle these components, we apply a set of three-body unitaries on the links of the square lattice as follows:
\begin{align}
&U=\prod_i U^x_i~ U^y_i,~\nonumber\\
    &U^x_{i}=\frac{(1+Z_i Z_{i+\hat{x}})}{2} \tilde{I}_{i+\frac{\hat{x}}{2}}+\frac{(1-Z_iZ_{i+\hat{x}})}{2} i\tilde{Y}_{i+\frac{\hat{x}}{2}},
    ~\nonumber\\
    &U^y_{i}=\frac{(1+Z_i Z_{i+\hat{y}})}{2} \tilde{I}_{i+\frac{\hat{y}}{2}}+\frac{(1-Z_iZ_{i+\hat{y}})}{2} i\tilde{Y}_{i+\frac{\hat{y}}{2}}    ~\nonumber\\
    &|\Psi_{SPT}\rangle=U~ \bigotimes_i \left(|\rightarrow \rangle_i \otimes |\tilde{\uparrow} \rangle_{i+\frac{\hat{x}}{2}} \otimes |\tilde{\uparrow} \rangle_{i+\frac{\hat{y}}{2}} \right)
    \label{unitary2d}
\end{align}
In this unitary, \( Z \) denotes the operator acting on the system qubits at the vertex, while \( \tilde{Y} \) operates on the ancilla located on the links. The symbols \( \hat{x} \) and \( \hat{y} \) represent the unit vectors in the square lattice. This three-body unitary operation, enforces the cluster of three spins on each link to have an even \( S_z \) parity.
\begin{align}
Z_i  \tilde{Z}_{i+\frac{\hat{y}}{2}} Z_{i+\hat{y}}=1,~~Z_i\tilde{Z}_{i+\frac{\hat{x}}{2}}Z_{i+\hat{x}} =1,
\end{align}

\begin{figure}[h!]
\includegraphics[width=0.46\textwidth]{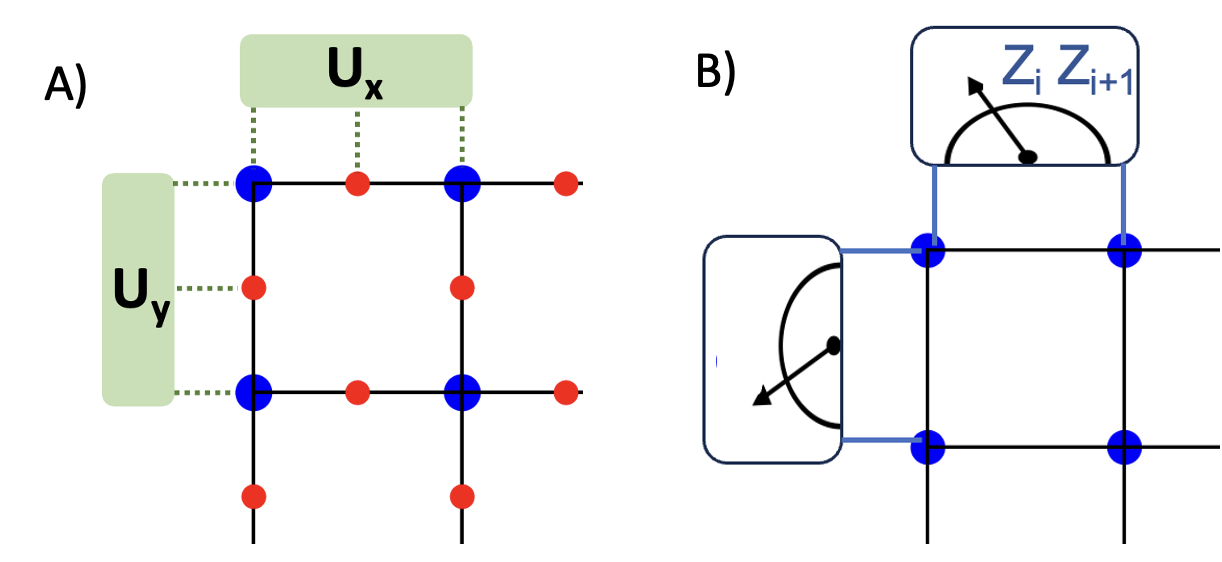}
\caption{\textbf{$2$D SSSB and purification from $1$-form SPT.} A) The unitary gates comprise a cluster of three-body gates acting on both the x-link and y-link.
B) Such a unitary in the extended Hilbert space is tantamount to quantum channels that measure the spin bilinears on the link.}
\label{2d}
\end{figure}

The unitary gates in Eq.~\ref{unitary2d} preserve the \( \mathbb{Z}_2 \) symmetry of the system qubits, denoted by \( X = \prod_i X_i \). After applying the unitary, the resultant wave function \(|\Psi_{SPT}\rangle \) represents an SPT state and is the ground state of the stabilizer Hamiltonian:
\begin{align}\label{2dsta}
H=-\sum_{b}\tilde{Z}_b \prod_{v \in e_b} Z_v -\sum_{s}X_s \prod_{b \in v_s} \tilde{X}_b
\end{align}
Here \( \tilde{Z}_b \) represents the ancilla qubit that resides on link $b$, and \( v \in e_b \) refers to the two vertex ends of the link where the system qubits are located. In the second term, \( X_s \) acts on the system qubits at the vertex, while \( a \in v_s \) spans the four links connecting to the vertex. This Hamiltonian is known as the parent Hamiltonian for the higher-form SPT protected by a 1-form \( \mathbb{Z}_2^A \) and a 0-form \( \mathbb{Z}_2^S \) symmetry\cite{verresen2022higgs}.
\begin{align}
\mathbb{Z}_2^A: \prod_{i \in \gamma} \tilde{Z}_i,~~
\mathbb{Z}_2^S: \prod_i X_i
\label{sym2d}
\end{align}
\( \mathbb{Z}_2^A \) represents the 1-form symmetry that acts on the ancillae, with \(\gamma\) denoting any closed loop along the links. \( \mathbb{Z}_2^S \) is the 0-form \( Z_2 \) symmetry (global symmetry) acting on the system qubits. Importantly, the unitary operator we apply preserves the \( \mathbb{Z}_2^S \) symmetry while simultaneously breaking the \( Z_2^A \) symmetry. This higher-form SPT state can be detected by a non-vanishing string order parameter\cite{verresen2022higgs,lee2022decoding,zhu2023nishimori}:
\begin{align}\label{string2d}
\langle O_s \rangle = \langle \prod_{j \in \partial l} Z_j (\prod_{i \in l} \tilde{Z}_{i} )  \rangle 
\end{align}
Here, \( l \) represents an arbitary open string along the link, and \( \partial l \) refers to the two endpoints of the string (on the vertex). The string order contains a product of the ancilla qubits \( \tilde{Z}_{i} \) along the string, decorated with two system qubits \( Z_j \) situated at the two ends of the string.
The presence of a non-vanishing string order indicates that the two distant system qubits share conditional mutual information mediated by the ancilla. The two-point correlation \( \langle Z_i Z_{j} \rangle \) is influenced by the 1-form charge of the ancilla located on the open string connecting them. As a result, projecting all ancillae into the \( |\tilde{\uparrow}\rangle \) state leads the system to evolve into a cat-state with long-range order\cite{piroli2021quantum,raussendorf2005long,verresen2021efficiently,tantivasadakarn2021long,verresen2022higgs}.

\begin{figure}[h!]
\includegraphics[width=0.48\textwidth]{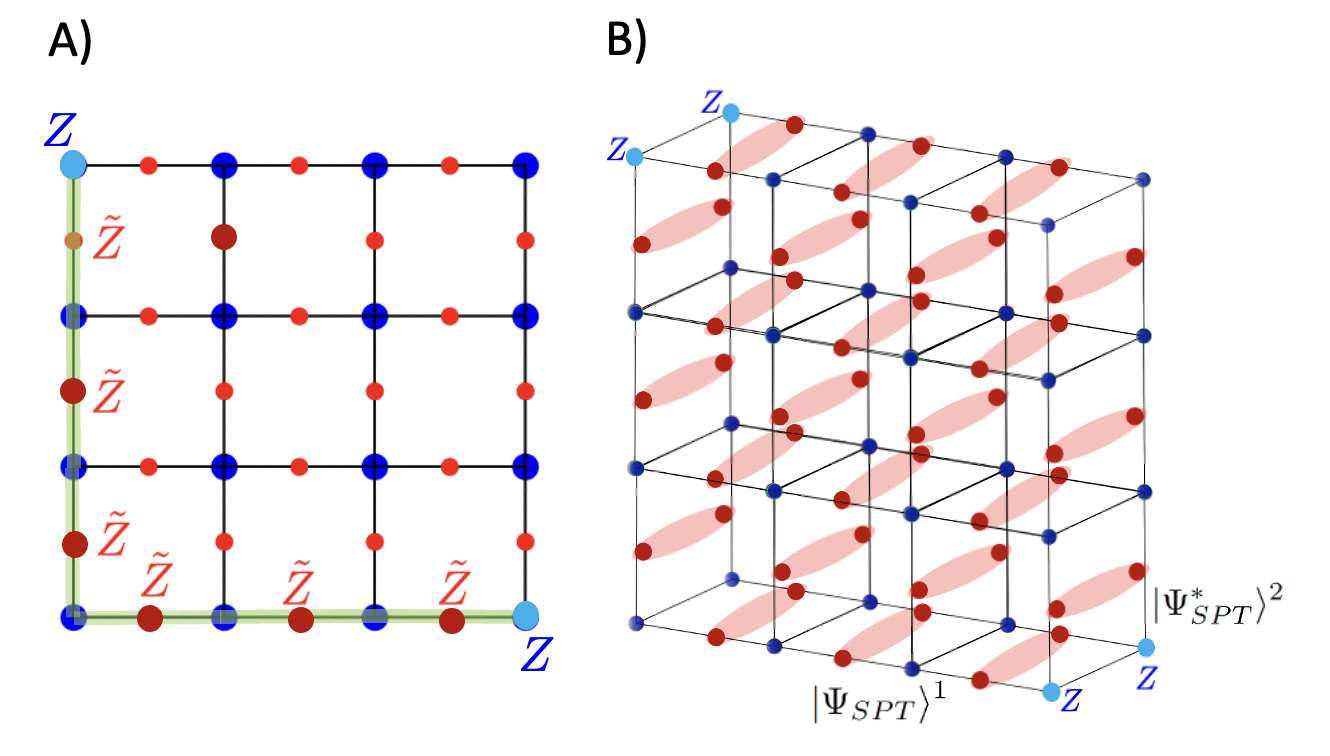}
\caption{\textbf{R{\'e}nyi-II correlator and EPR-induced long range order in $2$D.} A) The 1-form SPT exhibits a string order, where the two-point correlation (light blue dots) $\langle Z(x)Z (y)\rangle$ of the system qubits is influenced by the charge string of the ancilla (green string) $\prod_a \tilde{Z}_a$ that interconnects them.
B) Taking two copies of the SPT state \(|\Psi_{\text{SPT}}\rangle^{1} \otimes |\Psi_{\text{SPT}}\rangle^{2}\) and projecting the ancilla from both copies onto a symmetric EPR pair (illustrated by the red bond) results in long-range order (LRO) in the four-point correlator (represented by light blue dots).}
\label{double}
\end{figure}

Upon tracing over the ancilla from \( |\Psi_{SPT}\rangle \), we obtain a density matrix reminiscent of \( \rho_+ \) in Eq.~\ref{densitybreak}. The quantum channel corresponding to the unitaries in Eq.~\ref{unitary2d} is:
\begin{align}\label{decohere2d}
    & \hrho^D = \mathcal{E}[\hrho_0], \ \ \mathcal{E} = \prod_{\vect{i}} \cE^x_{\vect{i}}~\cE^y_{\vect{i}} , \nonumber \\ 
    &  \cE^x_{\vect{i}}[\hrho_0] = \frac{1}{2} \hrho_0 + \frac{1}{2} Z_i Z_{i+\hat{x}} \hrho_0 Z_i Z_{i+\hat{x}} \nonumber \\ 
     &  \cE^y_{\vect{i}}[\hrho_0] = \frac{1}{2} \hrho_0 + \frac{1}{2} Z_i Z_{i+\hat{y}} \hrho_0 Z_i Z_{i+\hat{y}} 
\end{align}
This quantum channel preserves the strong \( \mathbb{Z}_2 \) symmetry, defined as \( X = \prod_i X_i \), which indeed corresponds to the \( \mathbb{Z}_2^S \) symmetry defined for the purified state in Eq.~\ref{sym2d}.

By tracing out the ancilla qubits on the links, the mixed state density matrix exhibits spontaneous breaking of the strong \(\mathbb{Z}_2\) symmetry that can be characterized by the Rényi-2 correlator, which operates on the doubled density matrix as defined in Eq.~\ref{doubleco}.
As discussed in the previous section, this corresponds to the duplicate two-point correlation functions evaluated on the EPR-projected state $\ket{\psi}_{pp}$:
 \begin{align}\label{eq:renyi22d}
    &\frac{{\rm tr}\left(Z_i Z_j\hrho Z_i Z_j\hrho \right)}{{\rm tr}(\hrho^2) } =\langle \Psi |_{pp}
Z^1_i Z^2_i  Z^1_{j}Z^2_{j}|\Psi \rangle_{pp}.
\end{align}

Now we consider the general decoherence channel with a finite measurement rate:
\begin{align}\label{decohere2dweak}
    & \hrho^D = \mathcal{E}[\hrho_0], \ \ \mathcal{E} = \prod_{\vect{i}} \cE^x_{\vect{i}}~\cE^y_{\vect{i}} , \nonumber \\ 
    &  \cE^x_{\vect{i}}[\hrho_0] = (1-p) \hrho_0 + p Z_i Z_{i+\hat{x}} \hrho_0 Z_i Z_{i+\hat{x}} \nonumber \\ 
     &  \cE^y_{\vect{i}}[\hrho_0] = (1-p) \hrho_0 + p Z_i Z_{i+\hat{y}} \hrho_0 Z_i Z_{i+\hat{y}} 
\end{align}
which maintains strong \(\mathbb{Z}_2\) symmetry for any value of \(p\). 

Can spontaneous strong symmetry breaking now occur at a finite rate $p<1/2$? To investigate the impact of this general quantum channel, we return to the corresponding unitary operators acting on the enlarged Hilbert space. As in the 1D case we define a set of 3-body unitary gates that operate on the links:
\begin{align}\label{unitary2dweak}
&U(\theta)=\prod_i U^x_i(\theta)~ U^y_i(\theta),~\nonumber\\
    &U^x_{i}(\theta)=\frac{(1+Z_i Z_{i+\hat{x}})}{2} (\cos(\theta) \tilde{I}_{i+\frac{\hat{x}}{2}}+i\sin(\theta) \tilde{Y}_{i+\frac{\hat{x}}{2}})  ~\nonumber\\
    &+\frac{(1-Z_iZ_{i+\hat{x}})}{2} (\sin(\theta) \tilde{I}_{i+\frac{\hat{x}}{2}}+i\cos(\theta) \tilde{Y}_{i+\frac{\hat{x}}{2}})),
    ~\nonumber\\
    &U^y_{i}(\theta)=\frac{(1+Z_i Z_{i+\hat{y}})}{2} (\cos(\theta) \tilde{I}_{i+\frac{\hat{y}}{2}}+i\sin(\theta) \tilde{Y}_{i+\frac{\hat{y}}{2}})  ~\nonumber\\
    &+\frac{(1-Z_iZ_{i+\hat{y}})}{2} (\sin(\theta) \tilde{I}_{i+\frac{\hat{y}}{2}}+i\cos(\theta) \tilde{Y}_{i+\frac{\hat{y}}{2}}))    ~\nonumber\\
    &|\Psi_{SPT}(\theta)\rangle=U(\theta)~ \prod_i |\rightarrow \rangle_i \otimes |\tilde{\uparrow} \rangle_{i+\frac{\hat{x}}{2}} \otimes |\tilde{\uparrow} \rangle_{i+\frac{\hat{y}}{2}}
\end{align}
$\theta$ controls the measurement rate with $p=\frac{1-\sin(2\theta)}{2}$. 
When \(\theta=0\), the gate simplifies to the unitary introduced in Eq.~\ref{unitary2d}, initiating a quantum channel of pure measurement with \(p = \frac{1}{2}\). At \(\theta = \frac{\pi}{4}\), the unitary only rotates the ancilla qubits, so the system and ancilla remain entangled.

Upon applying \(U(\theta)\), the resulting wave function \(|\Psi_{SPT}(\theta)\rangle\) retains the \(\mathbb{Z}_2^A\) 1-form symmetry only at \(\theta=0\). For other values of \(\theta\), the wave function loses its 1-form symmetry, causing the string order defined in Eq.~\ref{stringdual} to vanish\cite{verresen2022higgs}. Despite this, several significant characteristics of the higher-form SPT persist even after the explicit symmetry breaking. As discussed in Ref.~\cite{verresen2022higgs,tantivasadakarn2023building,tantivasadakarn2023pivot,lu2023characterizing}, higher-form symmetries can manifest phenomena distinctly different from those of conventional 0-form global symmetry. For example, the ground state degeneracy resulting from the spontaneous symmetry breaking (SSB) of a 1-form symmetry (interpreted as topological degeneracy) remains robust against the explicit breaking of the 1-form symmetry. Similarly, if we weakly break the 1-form symmetry in an SPT state, the edge mode continues to be gapless up to a certain threshold\cite{su2023higher}. These unique attributes arise because 1-form symmetries, though not present at the ultraviolet (UV) level, can emerge at the infrared (IR) level in a gapped system with a finite correlation length\cite{hastings2005quasiadiabatic}. In our subsequent discussion, we will show that the purified state \(|\Psi_{SPT}(\theta)\rangle\) with weak 1-form symmetry breaking (up to a certain threshold), exhibits a non-vanishing strange correlator\cite{lu2020detecting}. This strange correlator precisely captures the mixed-state long-range order resulting from SSSB.

As is delineated in Appendix.~\ref{app:ising}, the Rényi-2 correlator is mapped to the thermal two-point spin-spin correlator of the 2D classical Ising model (as per Eq.~\eqref{eq:Renyi2_to_Ising}) at an inverse temperature of $2\beta$ with $\tanh(\beta)=p/(1-p)$. Consequently, there exists an extended region above a critical error rate $p_c=\frac{1}{2}(1-\sqrt{\sqrt{2}-1})\approx 0.178$\cite{bao2023mixed,chen2023symmetry} (equivalently, below a critical $\theta_c$ with $\sin(2\theta_c)=\sqrt{\sqrt{2}-1}$), where the Rényi-2 correlator remains finite. This represents a spontaneous strong-to-weak symmetry-breaking transition triggered by local quantum channels in 2D.

\section{Averaged strange correlator in purified state}\label{sec:strange}

At this point, we have demonstrated that the mixed-state long-range order, characterizing strong symmetry breaking induced by local quantum channels, can be considered a subsystem of a purified SPT state in the extended Hilbert space. In this purification framework, the system's qubits are entangled with the ancilla through local unitary gates. They exhibit conditional long-range mutual information, contingent on the ancilla's projection (or relatedly, conditional long-range order as measured by two-point correlators). The Rényi-2 correlator, pertinent to SSSB, can be manifested by projecting the ancillae in two copies of the SPT state into an EPR state and then measuring the post-projection state's correlation function. This mapping reveals that the SSSB in the mixed-state can be traced back to the measurement-induced long-range order in the purified state. Upon projecting the ancilla, the system's qubits in the post-projection state acquire long-range order.

To render this mapping precise, a crucial question on our agenda involves establishing a one-to-one correspondence between physical observables in the mixed state and their purified counterpart. In Sec.~\ref{sec:2d}, we examined the spontaneous strong $\mathbb{Z}_2$ symmetry breaking in 2D, triggered by quantum channels with a measurement rate $p$, and determined that the SSSB transition can occur at a specific finite rate $p_c$. Notably, these quantum channels can be represented as local unitary operations acting on the purified state in the extended Hilbert space. The purified wave function is proximate to the SPT wave function when $p\leq 1/2$ with weak 1-form symmetry breaking. A natural question arises: Can we identify any parameter in the purified state that exhibits singularity at the $p_c$ threshold?

This question involves two subtleties: 1) When \( p < 1/2 \), the corresponding purified state \( |\Psi(\theta)\rangle \) (where \( \theta > 0 \)) explicitly breaks 1-form symmetry, rendering the `SPT' nature ill-defined. 2) Varying the measurement rate \( p \) for the quantum channel defined in Eq.~\ref{decohere2dweak} corresponds to varying \( \theta \) for the local unitary gates in the purification framework, as specified in Eq.~\ref{unitary2dweak}. This implies that, although the Rényi-2 correlator (which is non-linear in the mixed density matrix \( \hat{\rho} \)) captures strong symmetry breaking with a singularity at the critical value \( p_c \), the purified state \( |\Psi(\theta)\rangle \) change smoothly by varying \( \theta \). The critical question then arises: How can we identify the decoherence-induced phase transition in a mixed state from the perspective of purification? 

Ref.~\cite{you2014wave,lepori2023strange,zhang2022strange} proposed the `strange correlator' as an efficient measure to probe SPT order:
\begin{align}\label{sc}
O^s(x,y)=\frac{\langle \Psi_{trivial} | \hat{O}^{\dagger}(x) \hat{O}(y) | \Psi_{SPT} \rangle}{\langle \Psi_{trivial} | \Psi_{SPT} \rangle}.
\end{align}
Specifically, the strange correlator applies a pair of distant charged operators $\hat{O}^{\dagger}(x) \hat{O}(y)$ (which could be chosen to be $Z(x) Z(y)$ for $\mathbb{Z}_2$ symmetry) to the SPT wave function and takes the inner product with a trivial wave function, for example, a symmetric product state. The strange correlation, notably, can exhibit long-range order for both the SPT state and the symmetry breaking state. Ref.~\cite{you2014wave} elucidated that the strange correlator can be mapped to the correlation function of a higher-dimensional Wess-Zumino-Witten theory, which exhibits either long-range or quasi-long-range order.

As we demonstrated in Sec.~\ref{sec:2d}, a mixed state exhibiting spontaneous strong symmetry breaking in 2D corresponds to the purified state \(|\Psi(\theta)\rangle\), which weakly breaks 1-form symmetry. Consequently, the salient features of mixed-state SSSB should be traceable from the purified wave function. To leverage this interrelation, we analyze the \textit{disorder-averaged strange correlator} of \(|\Psi(\theta)\rangle\). We will demonstrate that the non-vanishing disorder-averaged strange correlator in the purified state precisely captures the SSSB in the mixed state.

We begin by examining the strange correlator of the 2D purified wave function in Eq.~\ref{unitary2dweak},
\begin{widetext}
\begin{align}\label{puresc}
   &  O^s(x,y)=\frac{\langle \Psi_{trivial} | Z(x) Z(y) | \Psi_{SPT}(\theta) \rangle}{\langle \Psi_{trivial} | \Psi_{SPT}(\theta)\rangle},~~~~\ket{\Psi_{trivial}(s_{ij})}=~ \otimes_{\langle i,j \rangle}|s_{ij}=1  \rangle \otimes_i | x_i=1 \rangle \nonumber\\
   &\ket{\Psi_{SPT}(\theta)}=\sum_{\{ z_i \}} \prod_{\langle i,j\rangle} \left( \frac{1+z_iz_j}{2}(\cos{\theta}|s_{ij}=1\rangle -\sin{\theta}|s_{ij}=-1\rangle )\right.
\left.+\frac{1-z_iz_j}{2}(\sin{\theta}|s_{ij}=1\rangle -\cos{\theta}|s_{ij}=-1\rangle ) \right)  |\{ z_i \}\rangle 
\end{align}
\end{widetext}
In this context, \( s_{ij} \) represents ancilla qubits (in the \( Z \) basis), situated on the links of the square lattice between the nearest vertices at sites \( i \) and \( j \). The term \( z_i (x_i) \) denotes the spin pattern of the system qubits at vertex $i$ in the \( Z(X) \) basis. $|\{ z_i \}\rangle$ refers to a specific many-body pattern and we sum over all possible patterns.
For \( \theta = 0 \), \( \ket{\Psi_{SPT}(0)} \) corresponds to the exact SPT state with 1-form \( \mathbb{Z}_2^A \) symmetry (acting on the ancilla) and 0-form \( \mathbb{Z}_2^S \) symmetry (acting on the system), as defined in Eq.~\ref{sym2d}. When \( \theta \) is nonzero, the 1-form \( \mathbb{Z}_2^A \) symmetry is broken explicitly. At \( \theta = \pi/4 \), the ancilla spins decouple from the system qubits. The trivial wave function \( |\Psi_{trivial}\rangle \) is symmetrically chosen, with the ancilla polarized in the \( S^z \) direction as \( |s_{ij} = 1\rangle \) and the system qubits polarized in the \( S^x \) direction as \( |x_i = 1\rangle \).
After some derivation(see Appendix.~\ref{app:sc}), we can show that the strange correlator in Eq.~\ref{puresc} corresponds to the spin-spin correlator of the 2D Ising model at an effective temperature given by \(\tilde{\beta} =- \frac{\ln(\tan(\theta))}{2}\).

\subsection{Annealed disorder average strange correlator and type-II strange correlator}

Once we trace out the ancilla, the information of the ancilla qubits is lost, resulting in an averaging out of their information. We can envision this process as projecting each ancilla $s_{ij}$ into different patterns and then taking the disorder average for the post-projected wave function. In the context of the strange correlator, we consider measuring the \textit{annealed disorder-averaged strange correlator} by choosing various trivial states with a random assortment of ancilla spins \( \{ s_{ij}=\pm 1\} \), and then \emph{uniformly} averaging over these random selections.
\begin{align}\label{annealed}
   & \langle \overline{O^s(x,y)} \rangle ^2=\frac{\sum_{\{ s_{ij} \}}|\langle \Psi_{trivial}(\{ s_{ij} \}) | Z(x) Z(y) | \Psi_{SPT} \rangle|^2}{\sum_{\{ s_{ij} \}}|\langle \Psi_{trivial}(\{ s_{ij} \}) | \Psi_{SPT} \rangle|^2}\nonumber\\
 &= \frac{\sum_{\{ s_{ij} \}}\sum_{\{ z^1_i, z^2_i \}}(z^1_x z^1_y z^2_x z^2_y)e^{\tilde{\beta} \sum_{\langle i,j\rangle}s_{ij}(z^1_i z^1_j+z^2_i z^2_j)}}{\sum_{\{ s_{ij} \}} \sum_{\{ z^1_i, z^2_i \}}e^{\tilde{\beta} \sum_{\langle i,j\rangle}s_{ij}(z^1_i z^1_j+z^2_i z^2_j)}}
\end{align}
 The trivial state $\ket{\Psi_{trivial}(\{ s_{ij} \})}$ is a direct product of system spins polarized in the $S^x$ direction and a random assortment of ancilla spins \( \{ s_{ij}=\pm 1\} \) polarized in the $\pm S^z$ directions. Here, we average over all \( \{ s_{ij}=\pm 1\} \) with the same probability. 
Eq.~\ref{annealed} is akin to the \textit{annealed average} of the two-point correlation function in the random bond Ising model (RBIM) along the Nishimori Line\cite{nishimori1981internal}, 
which corresponds to the thermal correlation function evaluated on the $2D$ Ising model at finite temperature $\beta$ (with $\tanh(\beta)=(1-\sin(2\theta))/(1+\sin(2\theta))$ agreeing with that of the R{\'e}nyi-$2$ correlator. See Appendix.~\ref{app:sc} for detailed derivations).

Surprisingly, the annealed averaged strange correlator of the purified state \(|\Psi_{SPT}(\theta)\rangle\) is dual to the type-II strange correlator of the mixed state after the ancilla has been traced out. While the type-II strange correlator was originally proposed for decoherent SPT mixed states in Ref.~\cite{LeeYouXu2022}, our findings reveal that it stems from the annealed-average strange correlator in the enlarged Hilbert space resulting from purification. 
Given that the operators \(Z(x) Z(y)\) in the strange correlator act only on the system's qubits, we can alternatively express Eq.~\ref{annealed} as:
\begin{align}\label{type2}
    &\langle \overline{O^s(x,y)} \rangle ^2=\frac{\sum_{\{ s_{ij} \}} |\langle \Psi_{trivial}(\{ s_{ij} \}) | Z(x) Z(y) | \Psi_{SPT}(\theta)\rangle|^2}{\sum_{\{ s_{ij} \}} |\langle \Psi_{trivial}(\{ s_{ij} \}) | \Psi_{SPT}(\theta)\rangle|^2}\nonumber\\
     &= \frac{\sum_{\{ s_{ij} \}} |\langle s_{ij}| \otimes  \langle x_i=1| Z(x) Z(y) | \Psi_{SPT}(\theta) \rangle|^2}{\sum_{\{ s_{ij} \}} |\langle s_{ij}| \otimes  \langle x_i=1| \Psi_{SPT}(\theta)\rangle|^2}\nonumber\\
       &= \frac{\sum_{\{ s_{ij} \}}  |\langle x_i=1| Z(x) Z(y) \langle s_{ij}| \Psi_{SPT}(\theta) \rangle|^2}{\sum_{\{ s_{ij} \}} \langle x_i=1 |\langle s_{ij}|\Psi_{SPT}(\theta)\rangle|^2}\nonumber\\
       &=\frac{\Tr[\hrho^{0} Z(x) Z(y) \hrho Z(x) Z(y)] }{\Tr[\hrho^{0}  \hrho ]}
 \end{align}
The annealed average of the strange correlator in the purified state $|\Psi_{SPT}(\theta)\rangle$ exactly maps to the type-II strange correlator for the mixed state $\hrho$. Here, $\hrho^{0}$ is chosen to be a pure state density matrix consisting of a trivial tensor product state $|x_i=1\rangle$, and $\hrho$ refers to the density matrix of the system obtained after tracing out the ancilla from $|\Psi_{SPT}(\theta)\rangle$.

Finally, we present an alternative expression of Eq.~\ref{annealed} that reveals the physical interpretation of the annealed average strange correlator, which is also the type-II strange correlator:
\begin{align}
    &\langle \overline{O^s(x,y)} \rangle ^2= \frac{\sum_{\{ s_{ij} \}}  \langle x_i=1| Z(x) Z(y) |\langle s_{ij}|\Psi_{SPT}(\theta) \rangle|^2}{\sum_{\{ s_{ij} \}} \langle x_i=1 |\langle s_{ij}|\Psi_{SPT}(\theta)\rangle|^2}\nonumber\\
       &= \frac{\sum_{\{ s_{ij} \}}p(s_{ij})  \langle x_i=1| Z(x) Z(y) |\hat{P}(s_{ij}) \Psi_{SPT}(\theta) \rangle|^2}{\sum_{\{ s_{ij} \}}p(s_{ij}) \langle x_i=1 |\hat{P}(s_{ij}) \Psi_{SPT}(\theta)\rangle|^2}\nonumber\\
       &p(s_{ij})= |\langle s_{ij}| \Psi_{SPT}(\theta) \rangle|^2,\nonumber\\
       &|\hat{P}(s_{ij}) \Psi_{SPT}(\theta) \rangle=\frac{|s_{ij}\rangle  \langle  s_{ij}|\Psi_{SPT}(\theta) \rangle}{|\langle s_{ij}| \Psi_{SPT}(\theta) \rangle|}
       \end{align} 
The $\hat{P}(s_{ij}) |\Psi_{SPT}(\theta) \rangle$ represents the post-projection wave function, obtained after projecting the ancilla qubits into a specific $s_{ij}$ sector. Meanwhile, $p(s_{ij})$ denotes the probability of measuring the ancilla in $|\Psi_{SPT}(\theta)\rangle$ with the measurement outcome $s_{ij}$. Based on this decomposition, the annealed average of the strange correlator can be interpreted as follows: 1) Projecting $|\Psi_{SPT}(\theta)\rangle$ into a specific sector $\{s_{ij}\}$ and measuring the strange correlator for the post-projected state $\hat{P}(s_{ij})| \Psi_{SPT}(\theta)\rangle$. 2) Averaging over the strange correlator for different post-projected states $\hat{P}(s_{ij})|\Psi_{SPT}(\theta)\rangle$, each weighted according to the Born probability of the respective measurement outcome.

\subsection{Fidelity correlator and averaged strange correlator}

In Ref.~\cite{fidpaper}, the authors demonstrate that the intrinsic measure for the spontaneous transition from strong to weak symmetry breaking is the fidelity correlator $F^1(\hrho)$,
\begin{align}\label{fid}
    &F^1(\hrho)=\left(\Tr{\sqrt{\sqrt{\hrho}Z(x)Z(y) \hrho Z(x)Z(y) \sqrt{\hrho}}}\right)^2
\end{align}
If we regard \(\hat{\sigma} = Z(x)Z(y)\hrho Z(x)Z(y)\) as another density matrix denoted by \(\hat{\sigma} \), the above expression indeed represents the fidelity between \(\hat{\sigma} \) and \(\hrho\). In particular, Ref.~\cite{fidpaper} demonstrates that nonzero fidelity implies the mixed-state density matrix is symmetrically non-invertible, thus rendering the strong \( \mathbb{Z}_2 \) symmetry operation non-localizable. Therefore, it can serve as a measure to pinpoint the SSSB transition in mixed states. For the 2D transition from strong to weak \(\mathbb{Z}_2\) symmetry breaking of interest, this fidelity operator corresponds to the random bond Ising model along the Nishimori line\cite{fidpaper}.

In this section, we aim to establish a connection between the fidelity measure of strong symmetry breaking in mixed states and the strange correlator formalism within the context of purification.
 Uhlmann's theorem~\cite{uhlmann1976transition}  ensures that the fidelity between two mixed states, $\hrho$ and $\hat{\sigma}$, corresponds to the maximum overlap between purification states \(\Psi_{\rho}\) and \(\Psi_{\sigma}\). 
\begin{align}
    &F^1(\hrho)=\text{max}|\langle \Psi_{\rho}|\Psi_{\sigma} \rangle|^2
\end{align}
Now, label the optimal purified states for the matrices $\hat{\sigma} = Z(x)Z(y)\hrho Z(x)Z(y)$ and $\hrho$ as $\Psi_{\rho}$ and $\Psi_{\sigma}$, respectively. Given that $Z(x)Z(y)$ are local operators and the purified states only contain short-range correlations, the optimal purification,
which maximize $|\langle \Psi_{\rho}|\Psi_{\sigma} \rangle|$ has the form,
\begin{align}\label{relation}
    &|\Psi_{\sigma}\rangle =Z(x)Z(y) U^a |\Psi_{\rho}\rangle
\end{align}
Here, $U^a$ are unitary gates that \textit{only} act on the ancilla space according to the Schr\"odinger–HJW theorem.
We can rewrite the overlap of the optimal purified states as:
\begin{align}\label{fidpur}
    &F^1(\hrho)=|\langle \Psi_{\rho}|Z(x)Z(y) U^a|\Psi_{\rho}
    \rangle|^2\nonumber\\
   &=|\sum_{|k\rangle} \langle \Psi_{\rho}U^a|k \rangle \langle k |Z(x)Z(y)|\Psi_{\rho}
    \rangle|^2\nonumber\\
    &=|\sum_{k}p_k C_k(x,y)|^2  \nonumber\\
   &C_k(x,y)=\frac{\langle k |Z(x)Z(y)|\Psi_{\rho} \rangle}{\langle k |\Psi_\rho\rangle},~ p_k=\langle \Psi_{\rho}U^a|k \rangle  \langle k |\Psi_\rho\rangle
\end{align}
Here, $|k\rangle$ denotes a complete set of basis states for the purified state, which are symmetric product states. $C_k(x,y)$ represents the strange correlator for the purified state $\Psi_{\rho}$ relative to the trivial state $|k\rangle$.  

The variable \( p_k \) can be viewed as a specific distribution of `probability amplitude' that depends on \( U^a \). 
It is determined by the wavefunction structure of the optimal purified states \( \Psi_{\rho} \) and \( \Psi_{\sigma} \).
Consequently, the expression $\sum_{k}p_k C_k(x,y)$ can be interpreted as the probability-weighted average of the strange correlator, calculated by considering all possible trivial states.
Eq.~\ref{fidpur} suggests that the probability-weighted average of the strange correlator for the optimal purified state corresponds to the fidelity correlator defined in Eq.~\ref{fid}. This relationship highlights the intrinsic connection between the fidelity correlator in mixed states and strange correlators in purification states.

Let us outline a few essentials. In Eq.~\ref{fidpur}, averaging the strange correlator for the purified state involves a probability-weighted average over all trivial states by summing over the basis of product states for both the system and the ancilla. This is distinct from the annealed averaging defined in Eq.~\ref{annealed}, where a specific product state for the system is maintained while only the ancilla basis is summed over. Furthermore, this argument holds only if we identify the exact form of the optimal purifications $\Psi_{\rho}$ and $\Psi_{\sigma}$, since the `probability amplitude' $p_k$ involves the $U^a$ operator in Equation~\ref{relation}. Indeed, we can treat $p_k=\langle \Psi_{\rho} U^a | k \rangle \langle k | \Psi_\rho \rangle$ as the scattering amplitude that transfers between $| U^a \Psi_{\rho} \rangle$ and $| \Psi_{\rho} \rangle$, mediated by the intermediate state $| k \rangle$.
In our previous 1D example in Eq.~\ref{unitary2} with parameters \(\theta=0\) and \(p=1/2\), \(U^a\) can be selected as a charge string of \(X\) acting on the ancilla between \(x\) and \(y\) to achieve optimal purification. For two general mixed-state density matrices, determining and implementing their optimal purification presents considerable challenges. Since the precise form of \(U^a\) is unknown, the probability distribution \(p_k\) also remains undetermined. Consequently, our findings reveal the interrelationship between the average strange correlator and fidelity correlator, yet the specific nature of this averaging process continues to be unclear.

\subsection{Measurement induced long-range order from purification}\label{sec:mie}
To summarize, we bridge the connection between SSSB in the mixed state and its SPT counterpart within the purified framework, and establish the correspondence between physical observables from both sides. In particular, we demonstrate that measurement-induced long-range order, along with the annealed-averaged strange correlator in the purified state, can be dual to the Rényi-2 correlator and type-II strange correlator in the mixed state. Likewise, for optimal purification, the `probability-averaged' strange correlator maps to the fidelity correlator of the mixed state. Below is a table that summarizes their correspondence.


\begin{widetext}
\begin{center}
\begin{table}[!htbp]
\fontsize{12}{12}
\begin{tabular}{| l | l |}
\hline
 \textbf{Purified state} $\bm{\Psi}$ & \textbf{Mixed state} $\bm{\hrho}$ \\ \hline   
 \vtop{\hbox{\strut EPR-projection induced correlation:}\hbox{\strut $\langle \Psi |_{pp}
Z^1(x) Z^2(x)  Z^1(y) Z^2(y)|\Psi \rangle_{pp}$}} &  \vtop{\hbox{\strut Rényi-2 correlator:}\hbox{\strut $\frac{\Tr \left(Z(x) Z(y)\hrho Z(x) Z(y)\hrho \right)}{\Tr \hrho^2}$}} \\ \hline \vtop{\hbox{\strut Annealed-averaged strange correlator:}\hbox{\strut $\frac{\sum_{\{ s_{ij} \}} |\langle \Psi(\{ s_{ij} \}) | Z(x) Z(y) | \Psi_{SPT} \rangle|^2}{\sum_{\{ s_{ij} \}} |\langle \Psi(\{ s_{ij} \}) | \Psi_{SPT} \rangle|^2}$}} &  \vtop{\hbox{\strut Type-II strange correlator:}\hbox{\strut $\frac{\Tr[\hrho^{0} Z(x) Z(y) \hrho Z(x) Z(y)] }{\Tr[\hrho^{0}  \hrho ]}$}}  \\ \hline
 \vtop{\hbox{\strut Averaged strange correlator for optimal purification:}\hbox{\strut $|\sum_{|k\rangle} \langle \Psi_{\rho}U^a|k \rangle \langle k |Z(x)Z(y)|\Psi_{\rho}
    \rangle|^2$}} & \vtop{\hbox{\strut Fidelity correlator:}\hbox{\strut $\left(\Tr{\sqrt{\sqrt{\hrho}Z(x)Z(y) \hrho Z(x)Z(y) \sqrt{\hrho}}}\right)^2$ }}  \\ \hline
\end{tabular}
\caption{Summary of the mapping between observables for mixed states and their corresponding purification. }
\end{table}
\end{center}
\end{widetext}

The strange correlator has an important connection to the phenomenon of measurement-induced mutual information\cite{lin2023probing}: in particular, for a general pure state on a composite system $ABCD$, any strange correlator between points $x \in A$ and $y \in B$ lower-bounds the mutual information between $A$ and $B$ generated by measuring all the spins in $D$ (which we consider to be the ancilla). The precise bound is derived in Ref.~\cite{cheng2023universal}. Measurement-induced mutual information is a phenomenon whereby measuring the degrees of freedom in some part of a system teleports information across that part, creating long-range correlations that did not previously exist. It is a characteristic property of SPT phases—indeed, it is what renders them useful as resource states for measurement-based quantum computation. In Appendix-\ref{app:mi}, we demonstrate that the annealed average of the strange correlator provides a lower bound for the measurement-induced mutual information.

From the examples we demonstrated in the aforementioned discussion, if the density matrix $\hrho$ has SSSB regarding strong symmetry $G_S$, the purified state $\Psi_{\rho}$ could be chosen to be an SPT state with $G_S \times G_A$ symmetry. Here, $G_S$ acts only on the system qubits (thus being the strong symmetry for the density matrix), and $G_A$ acts only on the ancilla. In particular, the SPT state exhibits a decorated domain wall structure between the $G_S$ and $G_A$ symmetries. Therefore, projecting the ancilla to a specific $G_A$ sector can trigger long-range order (or quasi-long-range order) for the system qubits\cite{lee2022decoding,zhu2023nishimori}.
However, since we are focusing solely on the strong to weak symmetry-breaking for the mixed-state, with the ancilla degrees of freedom being traced out, the $G_A$ symmetry acting on the ancilla is absent in the mixed-state density matrix. Likewise, as the purified state is not unique, other purifications exist, and the purified state does not necessarily need to exhibit $G_A$ symmetry in the ancilla. This is also reflected in our 2D example of strong $\mathbb{Z}_2$ symmetry-breaking discussion in Sec.~\ref{sec:2d}, where the purified state, despite breaking 1-form symmetry of the ancilla, still supports strong to weak $\mathbb{Z}_2$ symmetry breaking in the mixed state after tracing out the ancilla. 

Hence, we clarify that purification resulting in an SPT state is a sufficient but not necessary condition for strong symmetry breaking in a mixed state. As demonstrated in Sec.~\ref{sec:strange}, the strong symmetry breaking in the mixed state can be attributed to the non-vanishing averaged-strange correlator in the purified state. Thus, a transition from strong to weak symmetry breaking implies that the purified wavefunction contains measurement-induced long-range order: Upon projecting the ancilla, the post-projected state exhibits a long-range correlation. Such measurement-induced long-range order does not, in general, require the purified state to be an SPT state, although an SPT wavefunction with a decorated domain wall structure can be a prominent example of measurement-induced long-range order~\cite{verresen2021efficiently}.

\section{Generalized strong symmetry breaking driven by local quantum channels}\label{sec:generalsym}

\subsection{SSSB for 1-form symmetry in 2D: Mixed state topological order}

Understanding topologically ordered states in open quantum systems is both conceptually and practically important at the intersection of condensed matter and quantum information science. In this section, we illustrate how local quantum channels can potentially trigger a product state to evolve into a topologically ordered mixed state through the lens of SSSB of a strong 1-form \(\mathbb{Z}_2\) symmetry\cite{tantivasadakarn2023hierarchy,zhu2023nishimori,lee2022decoding}. In particular, we will demonstrate that the quantum channels driving the transition from strong to weak 1-form symmetry breaking can be mapped to a local unitary circuit in the extended Hilbert space. This circuit creates a symmetry-protected topological (SPT) state, protected by 1-form \(\mathbb{Z}_2\) symmetry (acting on the system) and global \(\mathbb{Z}_2\) symmetry (acting on the ancilla). The SPT wavefunction displays a membrane order parameter, which, upon integrating out the ancilla, corresponds to the correlation function of the 1-form \(\mathbb{Z}_2\) symmetry breaking.

\begin{figure}[h!]
\includegraphics[width=0.4\textwidth]{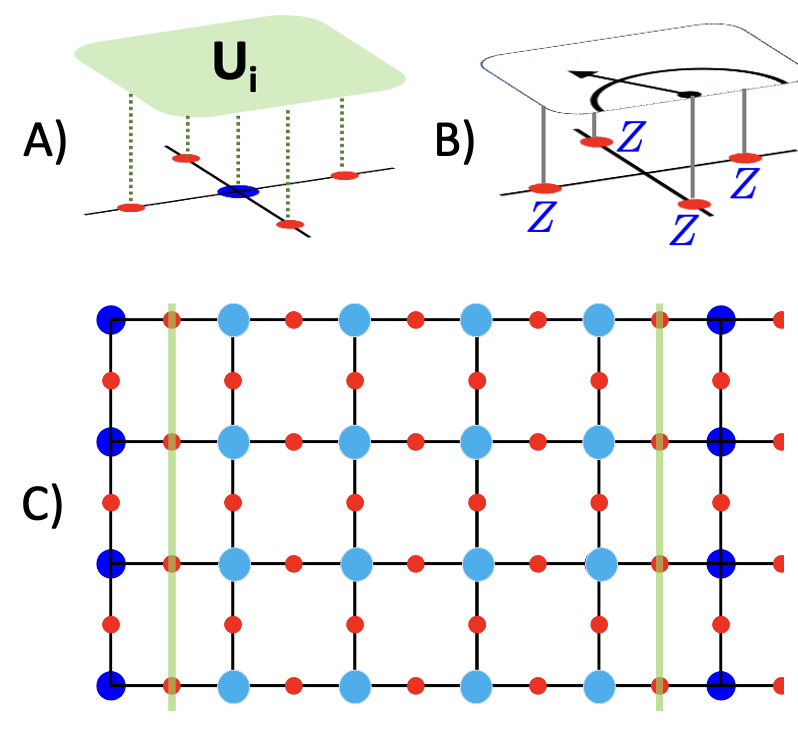}
\caption{\textbf{$2$D SSSB and mixed-state topological order.}  A) The unitary gates consist of a cluster of five-body gates acting on both the system qubits on the links (red) and the ancilla qubits on the vertices (blue).
B) A quantum channel measures the four system qubits (red) on the four links adjacent to the vertex. C) The higher-form SPT state can be identified by a non-vanishing membrane order parameter, where the correlation between the non-contractible Wilson loops (green lines acting on the system qubits) is locked by the total charge (light blue dots acting on the ancilla) situated between them.}
\label{to}
\end{figure}

We again begin with the purification perspective from the extended Hilbert space, which comprises the ancilla qubits located at the vertices, denoted as \( |\phi^A_0\rangle = \otimes_i |\tilde{\rightarrow} \rangle_i \) now polarized in \(S^x\), and system qubits living on the links, denoted as \( |\phi_0\rangle = \otimes_i |\uparrow \rangle_i \), polarized in \(S^z\). To entangle the system with the ancilla, we apply the following unitary:
\begin{align}
&U=\prod_i U_i,~\nonumber\\
    &U_{i}=\frac{1+\prod_{e \in v_i} X_e}{2} \tilde{I_i}+\frac{1-\prod_{e \in v_i} X_e}{2} \tilde{Z_i}
    \label{unitary1form}
\end{align}
In this context, \(X\) denotes the operator acting on the system qubit at each link, whereas \( \tilde{X} \) operates on the ancilla at the vertex. \( e \in v_i \) extends over the four links connecting to the vertex at site \(i\). This unitary gate is reminiscent of the \(\mathbb{Z}_2\) Gauss law, where the divergence of the system qubits at each vertex, denoted by \(\prod_{e \in v_i} X_e\), determines the ancilla charge \(\tilde{X}_i\) on the vertex. The unitary operation in Eq.~\ref{unitary1form} preserves the 1-form \(\mathbb{Z}^S_2\) symmetry of the system, given by $ G_S = \prod_{i \in \gamma} Z_i$, with \(\gamma\) representing any closed loop along the links.

The post-unitary state $|\Psi_{SPT}\rangle$ is the ground state of the following stabilizer Hamiltonian:
\begin{align}
H=-\sum_b Z_b \prod_{v \in e_b} \tilde{Z}_v -\sum_s \tilde{X}_s \prod_{b \in v_s} X_b
\end{align}
Here, \( b \in e_q \) refers to the two ends of the link where the ancilla qubits are located, while \( a \in v_s \) extends over the four links connecting to the vertex. This Hamiltonianshares the same form as the one discussed in Eq.~\ref{2dsta}, although the roles of system and ancilla qubits are interchanged. The ground state Hamiltonian is recognized as a higher-form SPT state, protected by a 1-form \( \mathbb{Z}_2^A \) symmetry acting on the system and a 0-form \( \mathbb{Z}_2^S \) symmetry acting on the ancilla.
\begin{align}
\mathbb{Z}_2^S: \prod_{i \in \gamma} Z_i,~~
\mathbb{Z}_2^A: \prod_i \tilde{X}_i
\label{sym2d1form}
\end{align} 
The higher-form SPT state \(|\Psi_{SPT}\rangle\) can be detected by a non-vanishing membrane order parameter:
\begin{align}
\langle O_s \rangle = \langle \Psi_{SPT} |\prod_{j \in \partial A} X_j ~(\prod_{i \in A} \tilde{X}_{i} ) | \Psi_{SPT}\rangle =1
\end{align}
Here, \(A\) represents a non-contractable area shaped like a half-cylinder, while \(\partial A\) refers to the two non-contractable loops at the left and right boundaries of area $A$. Such membrane order reproduces the \(\mathbb{Z}_2\) Gauss Law, stating that the total ancilla charge inside the area (\(\prod_{i \in A} \tilde{X}_{i}\)) is determined by the electric lines on the left and right boundaries of region $A$ as (\(\prod_{j \in \partial A} X_j\)). As these electric lines can be treated as the Wilson loop operator $W(x)=\prod_{j} X_{r+j \hat{y}}$, the membrane order indicates that the correlation of the Wilson loop operator \( \langle W(x_1) W(x_2) \rangle \) is determined by the total \(\mathbb{Z}_2^A\) charge of the ancilla between the two Wilson loops as Fig.~\ref{to}-C. 

Upon tracing out the ancilla, the system's density matrix, denoted as \(\hrho\), exhibits a mixed-state topological order with the spontaneous breaking of the strong 1-form \(\mathbb{Z}_2\) symmetry. The corresponding quantum channel associated with the unitary in Eq.~\ref{unitary1form} is:
\begin{align}
     &  \cE_{\vect{i}}[\hrho_0] = \frac{1}{2} \hrho_0 + \frac{1}{2} (\prod_{b \in v_s} X_b)  \hrho_0 (\prod_{b \in v_s} X_b). \nonumber\\
     & \hrho_0 =  |...\uparrow\uparrow\uparrow...\rangle \langle ...\uparrow\uparrow\uparrow...| , 
\end{align}
Such a quantum channel is akin to a pure measurement of four spins, \(\prod_{b \in v_s} X_b\), on every vertex.
The resulting spontaneous breaking of the strong 1-form \(Z_2\) symmetry can be measurable by the Rényi-2 correlator:
\begin{equation}\label{higherco}
    \frac{\Tr \left(W(x_1)W(x_2)\hrho W(x_1)W(x_2)\hrho \right)}{\Tr(\hrho^2)} =1
\end{equation}
As we have demonstrated, such a Rényi-2 correlator can be viewed as the correlation function of the post-projection wave function in the enlarged Hilbert space, which originates from duplicated copies of the purified wave function $\ket{\Psi_{SPT}}$, with ancilla copies projected into an EPR pair.

Now we consider a general decoherence channel driven by finite-rate measurement:
\begin{align}
     &  \cE_{s}[\hrho_0] = (1-p) \hrho_0 +p ~(\prod_{b \in v_s} X_b)  \hrho_0 (\prod_{b \in v_s} X_b). 
\end{align}
This quantum channel can be established by incorporating unitary gates as follows:
\begin{align}\label{unitary1formge}
    U_{s}(\theta)=&\frac{1+\prod_{b \in v_s}X_b}{2} (\cos(\theta)\tilde{I}_s+i\sin(\theta)\tilde{Z}_s)\nonumber\\
    +&\frac{1-\prod_{b \in v_s} X_b}{2} (-i\sin(\theta)\tilde{I}_s+\cos(\theta)\tilde{Z}_s)
\end{align}
When \(\theta=0\), the gate simplifies to the unitary introduced in \ref{unitary1form}. Conversely, at \(\theta=\pi/4\), the unitary only rotates the ancilla qubits, leaving the system qubits unaffected. Upon applying \(U(\theta)\) where $0<\theta<\frac{\pi}{4}$, the resultant state $|\Psi(\theta)\rangle$ breaks the global \(\mathbb{Z}_2^A\) symmetry, and the membrane order to vanish immediately\cite{lee2022decoding,zhu2023nishimori}. After tracing out the ancilla, the resultant density matrix, \(\hrho(\theta)\), evolves under the quantum channel as follows:
\begin{align}
    &  \cE_{\vect{s}}[\hrho_0] \nonumber\\
    &= (\frac{1+\sin(2\theta)}{2}) \hrho_0 + (\frac{1-\sin(2\theta)}{2}) (\prod_{b \in v_s} X_b) \hrho_0 (\prod_{b \in v_s} X_b). 
\end{align}
The mixed-state density matrix \(\hrho\) only exhibits spontaneous strong symmetry breaking (SSSB) for 1-form symmetry when \(\theta = 0\), indicating the absence of a strong 1-form symmetry-breaking transition at a finite measurement rate. Specifically, when computing the Rényi-2 correlation of Wilson operators in Eq.~\ref{higherco} for a general \(\hrho(\theta)\) where \(\theta>0\), it can be mapped onto the Wilson line correlator of the 2D \(\mathbb{Z}_2\) gauge theory at finite temperature, wherein charges are thermally excited\cite{lee2022decoding}. 
Given that the 2D \(\mathbb{Z}_2\) gauge theory does not demonstrate robust topological quantum memory\cite{castelnovo2007topological} with 1-form symmetry breaking at finite temperatures, the Rényi-2 correlation function in Eq.~\ref{higherco} vanishes for \(\theta \neq 0\) in the limit $|x_2-x_1|\to \infty$. From the SPT perspective within the extended Hilbert space, the purified state \(|\Psi(\theta)\rangle\) breaks the global \(\mathbb{Z}_2^A\) symmetry for \(\theta \neq 0\), and the strange correlator diminishes immediately. This is in stark contrast to the scenario discussed in Sec.~\ref{sec:2d}, where the SPT wavefunction, despite weak 1-form symmetry breaking, continues to exhibit non-vanishing strange correlators and measurement-induced long-range correlation.

\subsection{Subsystem SSSB in mixed state}
Before embarking into the generalization of strong-to-weak symmetry breaking to continuous symmetry groups, we conclude the study of discrete $\mathbb{Z}_2$ symmetry breaking in the context of subsystem symmetries. In particular, we consider a quantum channel on a 2D square lattice, where qubits are situated at the vertices and all are polarized in the x-direction, denoted as \(|\ldots\rightarrow\rightarrow\ldots\rangle\). A local quantum channel is applied, featuring a measurement channel that concurrently measures the four spins at the four corners of each plaquette, denoted as \(P\). 
\begin{align}
&  \cE_{P}[\hrho_0] = \frac{1}{2} \hrho_0 + \frac{1}{2} (\prod_{i \in P} Z_i)  \hrho_0 (\prod_{i \in P} Z_i).
    \label{subsystem}
\end{align}
These quantum channels display strong subsystem symmetry, conserving the \(\mathbb{Z}_2^{sub}\) charge on each row and column.
\begin{align}
    &\mathbb{Z}_2^{sub}: \prod_{i \in \text{row/column}} X_i
\end{align}

\begin{figure}[h!]
\includegraphics[width=0.46\textwidth]{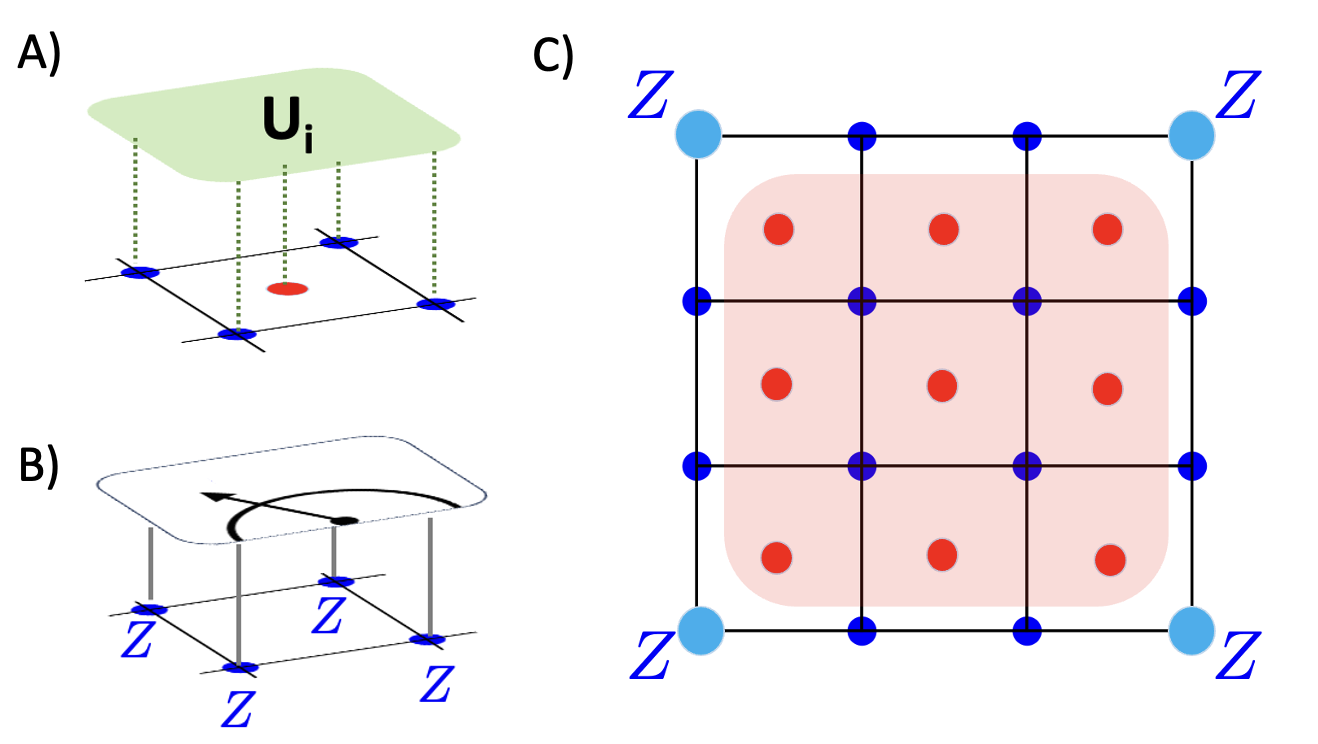}
\caption{\textbf{Subsystem SSSB in 2D.}  A) The unitary gates comprise a cluster of five-body gates on the plaquette.
B) A quantum channels that measure the four spins at the four corners of the plaquette. C) The subsystem SPT state can be identified by a non-vanishing membrane order parameter, where the four-point correlation (light blue dots) at the corner of the membrane is locked by the total charge (red) inside.}
\label{sspt}
\end{figure}

The density matrix of the post-quantum channel, denoted as \(\hrho\), exhibits long-range correlations that manifest the breaking of strong subsystem symmetry. This phenomenon is characterized by the four-point correlation function within the Rényi-2 correlator:
\begin{equation}\label{highercosub}
    \frac{\Tr \left(Z_{(0,0)} Z_{(x,0)} Z_{(0,y)} Z_{(x,y)}\hrho Z_{(0,0)} Z_{(x,0)} Z_{(0,y)} Z_{(x,y)} \hrho \right)}{\Tr (\hrho^2)} =1
\end{equation}
This four-point correlation function is reminiscent of the order parameter for 2D plaquette Ising models in the subsystem symmetry-breaking phase\cite{devakul2018classification,you2018subsystem}. The key difference here lies in our consideration of the spontaneous transition from strong to weak subsystem symmetry in a mixed state far from equilibrium. In this context, the charged operators within the correlation function must act on both the ket and bra spaces of the density matrix. Specifically, operators \(Z_{(0,0)} Z_{(x,0)}\) acting on the ket and bra of the density matrix are neutral under weak subsystem symmetry but are charged under strong subsystem symmetry along the y-columns. Consequently, the presence of a non-vanishing correlator indicates the breaking of strong subsystem symmetry.

Now, returning to the purification view, in the extended Hilbert space, the ancilla is positioned at the center of each plaquette. The initial state of the ancilla is a tensor product of spins polarized in the \(S_z\) direction. We apply the following unitary operator to each plaquette \(P\) to entangle the system with the ancilla:
\begin{align}
    U_{P}=\frac{(1+\prod_{i \in P} Z_i)}{2} \tilde{I}_P+\frac{(1-\prod_{ai \in P}Z_i)}{2} \tilde{X}_P
\end{align}
such that the global unitary is given by $U=\prod_P U_P$. Here, \(Z_i\) refers to the operator acting on the system qubits located at the vertices, while \(\tilde{X}_P\) acts on the ancilla situated at the center of each plaquette. \(\prod_{i \in P}\) includes four system spins positioned at the corners of the plaquette. Such a unitary operation exhibits a `decorated corner' feature: if the four spins at the four corners of a plaquette contain an odd number of \(\downarrow\) spins, we flip the ancilla at the center of plaquette to \(|\tilde{\downarrow} \rangle_P\).
The post-unitary wave function represents the ground state of the subsystem SPT Hamiltonian initially proposed in Ref.~\cite{devakul2018classification,you2018subsystem} with the following symmetries
\begin{align}
\mathbb{Z}_2^{sub,S}: \prod_{i \in \text{row/column}} X_i,~~
\mathbb{Z}_2^{sub,A}: \prod_{P \in \text{row/column}} \tilde{Z}_P
\end{align}
where \(\mathbb{Z}_2^S\) and \(\mathbb{Z}^A_2\) act on the system and the ancilla respectively. The unitary operator we apply preserves the \(\mathbb{Z}_2^S\) symmetry while breaking the \(\mathbb{Z}_2^A\) symmetry. The subsystem SPT state can be characterized by a non-vanishing membrane order parameter\cite{devakul2018classification,you2018subsystem}:
\begin{align}
O_s= \prod_{i \in C_A} Z_i (\prod_{a \in A} \tilde{Z}_{a} ) 
\end{align}
In this context, \(A\) denotes a rectangular-shaped membrane, and \(C_A\) represents the four corners of the membrane. This membrane order highlights that the four-point correlation, denoted as \(\langle \prod_{i \in C_A} Z_i \rangle\), is determined by the total \(S_z\) charge of the ancilla inside the membrane.
The Rényi-2 correlator in Eq.~\ref{highercosub} can be interpreted as the correlation function of the post-projection wave function in the enlarged Hilbert space.

Finally, we discuss the stability of strong subsystem symmetry breaking under general quantum channels. If the quantum channel in Eq.~\ref{subsystem} is replaced by a general quantum channel at a finite measurement rate \( p \), the corresponding purified state in the extended Hilbert space breaks the \( \mathbb{Z}_2^A \) symmetry, and its measurement-induced long-range order (LRO) is no longer present. Hence, strong subsystem symmetry breaking in 2D is unstable under a general quantum channel at a finite measurement rate \( p \). This conclusion is supported by the Rényi-2 correlator in Eq.~\ref{highercosub}, which maps to the 2D classical plaquette Ising model at finite temperature, where the subsystem symmetry breaking is absent due to strong thermal fluctuations\cite{xu2004strong}.
The situation would differ for 3D systems, where SSSB of subsystem symmetry can be driven by quantum channels at a finite measurement rate. We are eager to further investigate and expand upon this line of research in our future endeavors. 


\subsection{General SSSB in the mixed state for onsite symmetries}

The SSSB for $\mathbb{Z}_2$ symmetry can be generalized to more general
symmetries.
Here we summarize the results
for a global on-site symmetry with a finite symmetry group $G$,
and discuss their implications.
The reader is referred to Appendix~\ref{app:general} for details.

For a general finite symmetry group $G$,
we can construct the density matrix of the SSSB mixed state as
\begin{equation}
    \hrho_S \propto \sum_{g \in G} U_g ,
    \label{eq:SSSB-G}
\end{equation}
where $U_g = \prod_j u_g^{(j)}$ is the global symmetry transformation given by
the product of the local transformation $u_g^{(j)}$, which is
a representation of $g$ acting on the site $j$.
We introduce a multiplet of local order parameters $\calO^\alpha_\Omega$
defined on a local region $\Omega$.
We assume that they transform nontrivially under $G$, namely
${U_g}^\dagger \calO^\alpha_\Omega U_g = \sum_\beta M_{\alpha\beta}(g) \calO^\beta_\Omega$,
where $M_{\alpha\beta}(g)$ is a representation of $g \in G$ which does not contain
the identity representation.
Under this condition, the usual correlation function
between two distant regions $\Omega_1$ and $\Omega_2$
asymptotically vanishes in~\eqref{eq:SSSB-G}:
\begin{equation}
    \Tr{ \hrho_S \bar{\calO}^\alpha_{\Omega_1} \calO^\beta_{\Omega_2}  }  \to 0 ,
\end{equation}
when the distance between $\Omega_1$ and $\Omega_2$ is taken to $\infty$.

Nevertheless, the Rényi-2 correlation function
\begin{equation}
    \frac{\Tr{ \hrho_S \bar{\calO}^\alpha_{\Omega_1} \calO^\beta_{\Omega_2} \hrho_S
     \calO^\alpha_{\Omega_1} \bar{\calO}^\beta_{\Omega_2} } }{\Tr{(\hrho_S)^2}}
\end{equation}
does not vanish in the same limit, signaling the SSSB. In Sec.~\ref{sec:mie}, we demonstrated that a mixed-state density matrix with SSSB has a purified wave function with measurement-induced long-range order. A general protocol for finding the purification and corresponding FDLU state is called for in future studies.

\subsection{SSSB of continuous symmetry in mixed state?}

We briefly delve into the circumstances that may lead to spontaneous strong \(U(1)\) symmetry breaking in a mixed state, triggered by local quantum channels. Instead of focusing on quantum channels and the density matrix, we choose to conceptualize this process through purification in the extended Hilbert space generated by local unitary gates.

The extended Hilbert space consists of a system with an array of 1D fermion wires extended in the y-direction. Each wire contains two flavors of 1D fermions, whose left and right movers are denoted by \(\psi_{i,L} \) and \(\psi_{i,R}\) ($i=1,2$), respectively. \(\psi_{1}\) carries a unit charge and \(S_z = \frac{1}{2}\), while \(\psi_{2}\) carries a unit charge and \(S_z = \frac{-1}{2}\). The physical degrees of freedom in this enlarged Hilbert space encompass both \(U(1)\) charge and \(S_z\) spin (denoted \(U^s(1)\)). We will assume that the spins belong to the ancilla, while the charges belong to the system.
This setup resembles a bilayer fermion system with spins polarized in opposite directions. In this context, the fermion \(\psi_{i}\) represents a combination of degrees of freedom from both the system and the ancilla. The initial Hamiltonian of the system depicts a trivial atomic insulator, where the left and right fermions in each wire (with the same wire index \(x\)) are gapped by intra-wire coupling:
\begin{align}\label{h1}
   & H_1=\sum_{x}-\psi^{\dagger}_{1,L}(x)\psi_{2,R}(x)-\psi^{\dagger}_{2,L}(x)\psi_{1,R}(x)+h.c.
\end{align}
The initial state $|\Phi_0\rangle$ is invariant under charge U(1), but it breaks $U^s(1)$ symmetry. The ground state can be depicted as a trivial atomic insulator.

\begin{figure}[h!]
\includegraphics[width=0.4\textwidth]{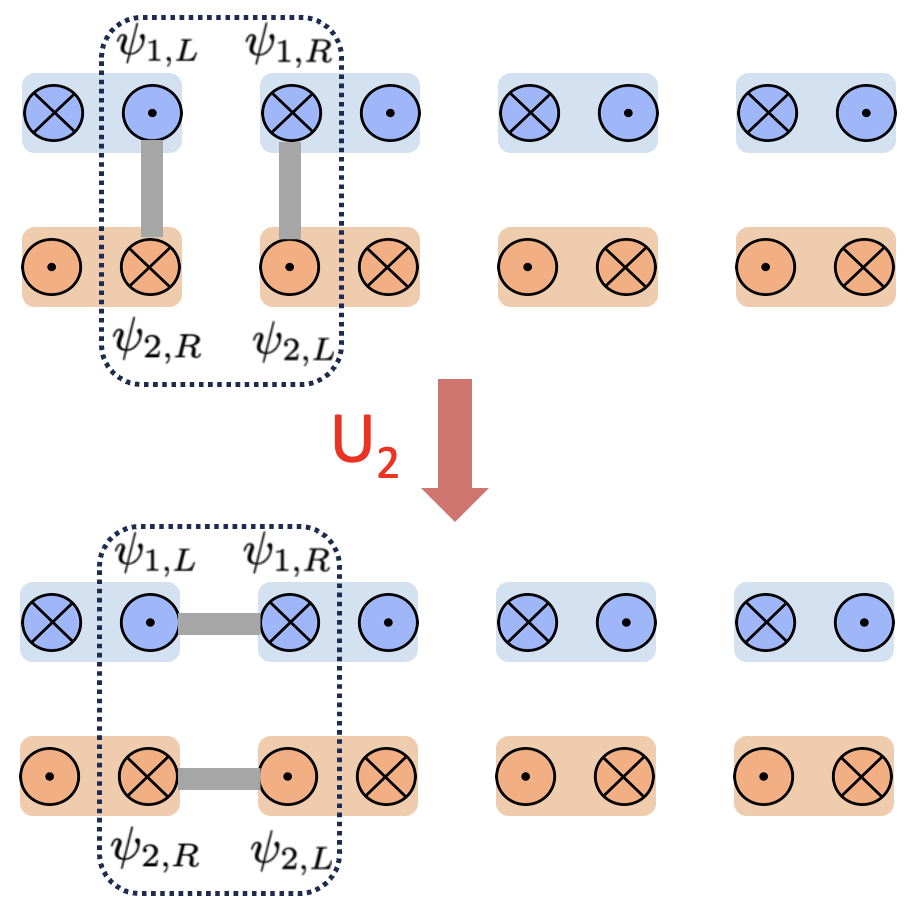}
\caption{The blue and orange unit cell represents fermionic 1D wires $\psi_{1,L/R}(x),\psi_{2,L/R}(x)$ extended along the y-direction. The dashed black box denotes a building block between two cells. After applying the unitary $U_2$ in each building block, the intra-wire coupling becomes the inter-wire coupling, which transforms the initial atomic insulator into a quantum spin Hall insulator.}
\label{qshgate}
\end{figure}

We now consider local unitary operations that transform our initial state $|\Phi_0\rangle$ into a quantum spin Hall-like state.
This unitary operator would be more transparent if we extract a building block as illustrated in Fig.~\ref{qshgate}. Each building block comprises four chiral fermions $(\psi_{2,R}(x), \psi_{1,L}(x), \psi_{2,L}(x+1), \psi_{1,R}(x+1))^T$ across two wires. The entire system is a tensor product of these building blocks. Within each building block, we apply a unitary operator as follows:
\begin{align}
      &U_2=\begin{pmatrix}
1 & 0 & 0 &0\\
0 & 0 & 1 &0\\
0 & 1 & 0 &0\\
0 & 0 & 0 &1
\end{pmatrix} \nonumber\\
   &U_2(\psi_{2,R}(x),\psi_{1,L}(x),\psi_{2,L}(x+1),\psi_{1,R}(x+1))^T\nonumber\\
   &=(\psi_{2,R}(x),\psi_{2,L}(x+1),\psi_{1,L}(x),\psi_{1,R}(x+1))^T
\end{align}
This unitary preserves the U(1) symmetry while breaking the $S_z$ conservation. The post-unitary Hamiltonian becomes:
\begin{align}\label{h1}
   & H_2=U^*_2 H_1 U_2 \nonumber\\
   &=-\sum_{x} \psi^{\dagger}_{2,R}(x)\psi_{2,L}(x+1)-\psi^{\dagger}_{1,L}(x)\psi_{1,R}(x+1)+h.c.
\end{align}
The $H_2$ Hamiltonian respects both U(1) symmetry and $S_z$ conservation, and is akin to the coupled-wire Hamiltonian for the quantum spin Hall effect\cite{kane2005quantum}, where spin-up and spin-down fermions form Chern bands with opposite Chern numbers. If we trace out the ancillae, the unitary operation in the extended Hilbert space manifests as a local quantum channel with strong \(U(1)\) symmetry, decohering the initial state \(\Phi_0\) into a mixed state density matrix \(\hrho\).

As demonstrated in Ref.~\cite{girvin1987off,hauser2023continuous,kvorning2020nonlocal,lu2023mixed,garratt2023measurements}, the quantum spin Hall state is characterized by string order with a power-law decay in correlation. This suggests that projecting the spin onto a specific pattern would induce charge quasi-long-range order. Upon tracing out the spin degree of freedom (ancillae), the resulting mixed state exhibits quasi-long-range order for the strong \(U(1)\) symmetry\cite{lu2023mixed}.
\begin{align}
 \frac{\Tr[e^{i \theta(r)} e^{-i \theta(r')} \hrho e^{-i \theta(r)} e^{i \theta(r')} \hrho ]}{\Tr[\hrho^2]} =\frac{1}{r^a}
\end{align}
Our results suggest that the local quantum channel can induce strong U(1) symmetry breaking accompanied by quasi-long-range order. Within the purification framework, the extended Hilbert space accommodates a quantum spin Hall-type state, characterized by a power-law decay in string order. Pressing questions arise, including identifying the precise Kraus operators resulting from the applied unitary gate within the quantum channel, and determining whether such U(1) symmetry-breaking transitions can be initiated by a general quantum channel with a finite \(p\). We will defer these inquiries to future studies.

\section{Implication for Mixed-state topological order and decoherence-induced transitions}\label{sec:topo}

\subsection{Noisy toric code from Wegner's duality}\label{sec:econ}

Topological quantum memory can protect information against local errors up to finite error thresholds. Recent studies have opened avenues for the characterization of mixed-state topological order triggered by local quantum channels\cite{lee2022decoding,LeeYouXu2022,zhu2023nishimori,chen2023separability}.
Refs.~\cite{bao2023mixed,fan2024diagnostics} investigate the noisy toric code by considering the ground state wave function \(|\Psi_{TC}\rangle\) of the 2D toric code with local phase errors, which can be quantitatively described as a quantum channel that measures the \(Z_i\) spin on each link:
\begin{align}\label{tc}
    & \hrho_0=|\Psi_{TC}\rangle \langle  \Psi_{TC}|,~\hrho^D = \mathcal{E}[\hrho_0], \nonumber \\ 
    &  \mathcal{E} = \prod_{\vect{i}} \cE_{\vect{i}},~~ \cE_{\vect{i}}[\hrho_0] =
    (1-p) \hrho_0 + p  Z_i \hrho_0 Z_i  
\end{align}
It has been shown that by tuning the measurement rate \(p\), the mixed state undergoes a transition that can be viewed as \(e\)-anyon condensation in the error-doubled field.

Inspired by our previous discussion on the correspondence between decoherence in the mixed state and unitary gates in the purification picture, we can map this quantum channel to a unitary operator in the extended Hilbert space. The extended Hilbert space includes the system of interest, initialized as the toric code ground state \(|\Psi_{TC}\rangle\), together with ancilla spins which also live on the links of the square lattice, initialized as \(|\phi^A_0\rangle = \bigotimes_i |\tilde{\uparrow}\rangle_i\). The unitary operator \(U\) entangles the system and ancilla spin on each link as follows:
\begin{align}
U(\theta)=&\prod_i U_i(\theta),~\nonumber\\
    U_{i}(\theta)=&\frac{(1+Z_i)}{2} (\cos(\theta) \tilde{I}_i+i\sin(\theta) \tilde{Y}_i)~\nonumber\\
    &+\frac{(1-Z_i)}{2} (\sin(\theta) \tilde{I}_i+i\cos(\theta)\tilde{Y}_i)  
\end{align}
The post-unitary state, upon tracing out the ancilla, reverts to the mixed-state density matrix induced by the quantum channel defined in Eq.~\ref{tc}, with an error rate \( p = \frac{1 - \sin(2\theta)}{2} \).

Suppose we apply the Wegner's (also known as Fradkin-Shenker) duality\cite{wegner1971duality,LeeYouXu2022,bao2023mixed,chen2023separability} to the initial toric code wave function \(|\Psi_{TC}\rangle\); the dual state becomes a paramagnetic state with qubits positioned at the vertices and polarized in \(S_x\) as \( |\phi_0\rangle = \bigotimes_i |\rightarrow\rangle_i \). The quantum channel in Eq.~\ref{tc} and its corresponding unitary in the purification picture exactly match the quantum channel introduced in Sec.~\ref{sec:2d}, which drives the transition from strong to weak \(\mathbb{Z}_2\) symmetry breaking. In particular, the Rényi-2 correlation that characterizes SSSB in Eq.~\eqref{eq:renyi22d} can be dual to the e-string operator, which characterizes the `anyon-condensation' transition in the decoherent toric code in error-doubled fields. Inspired by this duality, the SSSB in the mixed state is dual to the decoherent transition in the toric code, both of which can be reached through local quantum channels. 
Given that Wegner's duality is broadly applicable for relating quantum paramagnetic states with $G$ symmetry to the corresponding `gauged' lattice gauge theory, we anticipate that the correspondence between decoherent lattice gauge theory and SSSB can be extended to a broader class of scenarios, including fracton gauge theories and twisted gauge theories.

\subsection{Purification from the boundary of 3D TQFT}\label{sec:fcon}

Ref.\cite{sohal2024noisy,wang2023intrinsic} introduce an intrinsic mixed-state topological order by considering a 2-qubit quantum channel to the toric code ground state:
\begin{align}\label{tcf}
    & \hrho_0=|\Psi_{TC}\rangle \langle  \Psi_{TC}|,~\hrho^D = \mathcal{E}[\hrho_0], \nonumber \\ 
    &  \mathcal{E} = \prod_{\vect{i}} \cE_{\vect{i}}, ~\cE_{\vect{i}}[\hrho_0] =
    (1-p) \hrho_0 + p  Z_i X_{i+\hat{v}} \hrho_0 Z_i  X_{i+\hat{v}}
\end{align}
With $\hat{v}=(\frac{\hat{x}}{2},-\frac{\hat{y}}{2})$.
The two-body operator $Z_i X_{i+\hat{v}}$ creates an $f$-fermion in the toric code via the $e-m$ bound state. Thus, the mixed state undergoes a transition that can be viewed as $f\times \bar{f}$ -anyon condensation in the error-doubled field. Notably, an $f$-fermion condensate cannot be achieved in thermal equilibrium due to its fermionic statistics, and it becomes a unique and intrinsic feature in mixed-state quantum channels. In Ref.~\cite{sohal2024noisy}, the authors analyze such intrinsic anyon condensate from the Choi-double picture and elucidate its similarity with boundary topological order in 3+1d Walker-Wang models.

We aim to understand the topological mixed ensembles through the lens of the purification perspective. Reflecting on this approach, we highlight that certain topological mixed states can also be interpreted as the boundary density matrices of some 3D topological orders. 
Specifically, we consider the 3D topological order that has an open boundary at $z=0$, treating this as the purified state. The `system's degrees of freedom' are encoded in the top layer at $z=0$ with the rest being the ancilla. By tracing out the layers where $z<0$, we find that the reduced density matrix for the top layer closely resembles the mixed state density matrix specified in Eq.~\ref{tcf} or \ref{tc}, with $p=\frac{1}{2}$. 

\begin{figure}[h!]
\includegraphics[width=0.4\textwidth]{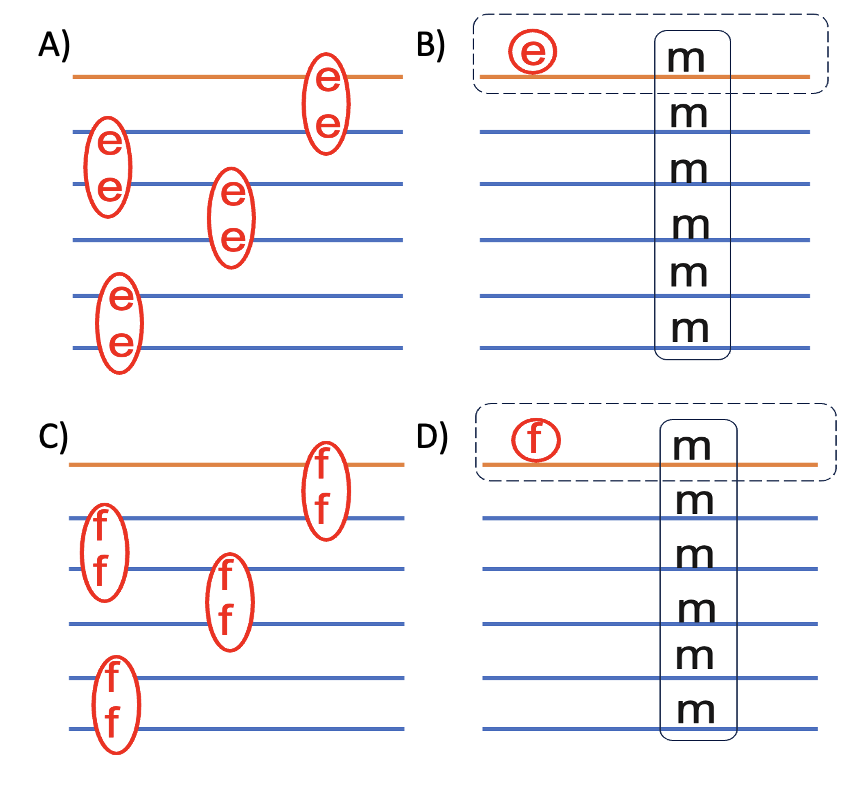}
\caption{ A-B) Stacks of toric code layers along z-direction and condense bound state of e-anyons between layers. The deconfined excitation is the e-anyon on each layer and the m-flux loop excitation. C-D) Stacks of toric code layers along z-direction and condense bound state of f-anyons between layers. The deconfined excitation is the f-anyon on each layer and the m-flux loop excitation. }
\label{surface}
\end{figure}

As delineated in Ref.~\cite{jian2014layer,Senthil_2015}, 3D topological order can be realized through coupled-layer constructions. This process is achieved by stacking layers of 2D topological order and by condensing bound states of anyons between these layers. For example, by starting with an array of 2D toric code layers aligned along the z-direction and condensing the bound state of e-anyons between the \(i\)-th and \((i+1)\)-th layers, the system transforms into a 3D \(\mathbb{Z}_2\) gauge theory. The condensation of e-anyon pairs between adjacent layers induces a coherence of charges across the layers, allowing a single e-excitation to remain deconfined and to move between layers within the bulk. Simultaneously, the isolated m-particles in each layer become confined. The resulting deconfined m-excitations form closed flux loops, as illustrated in Fig.~\ref{surface}. This structure corresponds to the flux loop excitation characteristic of 3D gauge theories. When there is an open surface at \(z=0\), this surface supports deconfined e-anyons. Flux excitations can terminate at this surface, resembling m-anyons; notably, these m-anyons on the surface are connected to the ends of flux strings extending into the bulk. By focusing on the top surface and tracing out the rest of the system, we are left with only the reduced density matrix \(\hrho^{top}\) as a mixed state for the \(z=0\) layer. It exhibits strong 1-form symmetry regarding the e-loop and weak 1-form symmetry regarding the m-loop. Consequently, the reduced density matrix for the top layer, \(\hrho^{top}\), resembles the decohered mixed state of the toric code with phase $Z$ error in Eq.~\ref{tc} with \(p=\frac{1}{2}\). The weak symmetry traces back to the connection between the m-anyons on the surface and the flux strings from the bulk. Here, the 3D topological order with an open surface is treated as the purification state, with all layers below (\(z<0\)) acting as ancillae being traced out.

Similarly, a different condensation channel involves condensing the bound state of f-fermions between the $i$-th and $i+1$-th layers. Since a pair of f-fermions constitutes a boson, enabling condensation in thermal equilibrium. The condensation of f-fermion pairs between adjacent layers induces coherence of fermion across the layers, allowing a single f-excitation to remain deconfined and to move between layers within the bulk. Simultaneously, the isolated m(e)-particles in each layer become confined. The resulting deconfined m(e)-excitations form closed flux loops(Fig.~\ref{surface}). At the open surface at z=0, it supports deconfined f-anyons. Flux excitations can terminate at the surface, resembling m-anyons on the surface; notably, these m-anyons are connected to the ends of flux strings extending into the bulk. Now, focusing on the top surface and tracing out the rest of the system, we are left only with the reduced density matrix, $\hrho^{top}$ for the z=0 layer. It exhibits strong 1-form symmetry regarding the f-loop and weak 1-form symmetry regarding the m-loop (e-loop). Thus, the reduced density matrix for the top layer $\hrho^{top}$ resembles the intrinsic topological mixed state with a two-bit error in Eq.~\ref{tcf} at $p=1/2$. While our approach does not reveal the unitaries that trigger the quantum channels, we have shown that the density matrix of some mixed-state topological orders can be visualized as the reduced density matrix of the top surface of 3D topological order.
A more careful analysis of this scenario will be considered in future work.

\section{Visualize decoherent SPT from purification perspective} \label{sec:SPT}

To this end, we have explored the spontaneous strong symmetry breaking from the purification perspective. Specifically, we demonstrated that the long-range order engendered by SSSB in a mixed state can be perceived as a subsystem of an SPT wave function within the extended Hilbert space. This space encompasses multiple degrees of freedom—denoted as the system of interest and ancilla qubits—that are highly entangled. Projecting onto the ancilla induces long-range correlations between the system qubits, eventually manifesting as the long-range order in the mixed state. In particular, the Rényi-2 correlator of the density matrix corresponds to the measurement-induced long-range order of the purified wave function in the doubled space\cite{ma2024symmetry}.

So far, it is clear that SSSB in the mixed state can be traced back to measurement-induced long-range order in the purified wave function. Therefore, it is natural to consider a wider variety of exotic mixed states in interacting systems, such as those exhibiting Symmetry-Protected Topological (SPT) orders in mixed states, induced by decoherence or quenched disorder\cite{LeeYouXu2022,ma2023topological,ma2023average, ma2024symmetry,de2022symmetry,chen2023symmetry,chen2023separability}. In Ref.~\cite{LeeYouXu2022,ma2023topological,ma2023average, ma2024symmetry,guo2024locally,xue2024tensor,deGroot2022}, several physical observables were proposed as diagnostics for the mixed-state SPT phase, including the Rényi-2 string order and the type-II strange correlator. In this section, we will provide a physical interpretation of these SPT observables from the purification perspective. While a detailed exploration of mixed-state SPT and its purification correspondence will be addressed in a forthcoming paper, we will briefly outline the key ideas here.

We begin with an SPT wave function protected by \(\mathbb{Z}_2^S\) and \(\mathbb{Z}_2^W\) symmetry introduced in Ref.~\cite{ma2023average}, which supports a non-vanishing string order:
\begin{align}\label{string1}
    O^s=\langle \Psi_{SPT}|Z_i \prod_{q \in l} X'_q Z_j|\Psi_{SPT}\rangle
\end{align}
Here, \(Z_i\)(\(Z_j\)) represents charge operators (carrying \(\mathbb{Z}_2^S\) charge) located at the ends of string \(l\), while \(\prod_{q \in l} X'_q\) acts as the \(\mathbb{Z}_2^W\) symmetry operator along string \(l\). The presence of non-vanishing string order suggests a mixed anomaly between \(\mathbb{Z}_2^W\) and \(\mathbb{Z}_2^S\).
As a point of interest, we consider that the SPT wave function encompasses multiple degrees of freedom, labeled as \(A\) and \(B\), which are extensive in space. These can be interpreted as degrees of freedom residing on different sublattices, layers, or different orbitals, among other possibilities. The symmetry operator \(\mathbb{Z}_2^S\) acts solely on \(A\), while both \(A\) and \(B\) are subject to \(\mathbb{Z}_2^W\) symmetry charge. Consequently, \(Z_i\)(\(Z_j\)) affects only \(A\), whereas the charge string \(\prod_{q \in l} X'_q\) impacts both \(A\) and \(B\).

If we treat \(B\) as the ancilla qubits from the environment and trace them out, the resulting reduced-density matrix, \(\hrho^A\), becomes a mixed-state SPT that possesses strong \(\mathbb{Z}_2^S\) symmetry and weak \(\mathbb{Z}_2^W\) symmetry. This strong symmetry arises because the purified state \(|\Psi_{SPT}\rangle\) does not permit the exchange of \(\mathbb{Z}_2^S\) symmetry charge between \(A\) and \(B\), ensuring that the subsystem \(A\) retains its conserved \(\mathbb{Z}_2^S\) charge. Conversely, since the \(\mathbb{Z}_2^W\) charge can fluctuate between subsystems \(A\) and \(B\), the reduced-density matrix \(\hrho^A\) exhibits only weak \(\mathbb{Z}_2^W\) symmetry.

If we calculate the string order parameter defined in Eq.~\ref{string1} based on the mixed state density matrix $\hrho^A$:
\begin{align}
  \Tr[\hrho^A (Z_i \prod_{q \in l_A} X_q Z_j)] \sim 0  
\end{align}
Here, the charge string \(\prod_{q \in l_A} X_q\) acts solely on qubits in subsystems \(A\). This string order vanishes because \(\mathbb{Z}_2^W\) represents a weak symmetry, allowing its charge to fluctuate between regions \(A\) and \(B\) in the purified wave function. Although the string order of the original purified state \(|\Psi_{SPT}\rangle\) in Eq.~\ref{string1} is non-vanishing, tracing out the ancilla (denoted as subsystem \(B\)) effectively averages out the even and odd charges carried by the ancilla along the string. Indeed, the decoherence effect obscures the string order due to the \(\mathbb{Z}_2^W\) charge fluctuation between the system and the ancilla.

To counteract the decoherence effect, we might consider creating two identical copies of both the system and the ancilla, denoted as \(|\Psi_{SPT}\rangle^{db} = |\Psi_{SPT}\rangle_1 \otimes |\Psi^*_{SPT}\rangle_2\).
We now project the two copies of the ancilla onto the EPR state, resulting in a post-projected wave function that exhibits non-vanishing string order.
\begin{align}\label{stringproject}
& \langle \Psi_{SPT}^{pp}|\prod_{t=1,2} (Z^t_i \prod_{q \in l_A} X^t_q Z^t_j)|\Psi_{SPT}^{pp}\rangle =\text{const} \nonumber\\
&|\Psi^{pp} \rangle = \frac{\hat{P}|\Psi^{db}_{SPT}\rangle}{|\langle \Psi^{db}_{SPT}|\hat{P}|\Psi^{db}_{SPT}\rangle|}
\end{align}
The projection of the double ancilla ensures that the total charge carried by the two ancilla copies (from system B) on the string is even. This condition allows the string order to be represented by a string operator acting solely on system \(A\). Here, \(Z_i,Z_j\) are operators that act only on \(A\), and \(\prod_{q \in l_A} X_q\) represents the \(\mathbb{Z}^W_2\) charge on the string, carried exclusively by system \(A\). Now, by transforming the second copy of the system and ancilla back into the bra space, the string order in Eq.~\ref{stringproject} becomes:
\begin{align}
&\langle \Psi_{SPT}^{pp}|\prod_{t=1,2} Z^t_i \prod_{q \in l_A} X^t_q Z^t_j|\Psi_{SPT}^{pp}\rangle   \rightarrow \frac{\Tr[O\hrho^A O \hrho^A]}{\Tr[(\hrho^A)^2]} \nonumber\\
   & O=Z_i \prod_{q \in l_A} X_q Z_j
\end{align}
This precisely maps to the Rényi-2 version of the string order.

Likewise, we can consider measuring the strange correlator in the purified state and subsequently taking its annealed average:
\begin{align}
    & \langle \overline{O^s(x,y)} \rangle ^2=\frac{\sum_{\{ x^B \}} |\langle \Psi_{trivial}(\{ x^B \}) | Z^A_i Z^A_j | \Psi_{SPT} \rangle|^2}{\sum_{\{ x^B \}} |\langle \Psi_{trivial}(\{ x^B \}) | \Psi_{SPT} \rangle|^2}\nonumber\\
     &= \frac{\sum_{\{ x^B \}} |\langle x^B | \otimes  \langle x^A_i=1| Z^A_i Z^A_j | \Psi_{SPT} \rangle|^2}{\sum_{\{ x^B \}} |\langle x^B| \otimes  \langle x^A_i=1| \Psi_{SPT}\rangle|^2}\nonumber\\
       &=\frac{\Tr[\hrho^{0} Z_i Z_j\hrho^A Z_i Z_j] }{\Tr[\hrho^{0}  \hrho^A ]}\nonumber\\
       &\hrho^{0}= |x^A_i=1\rangle \langle x^A_i=1 |,~ \hrho=\Tr_{B}( |\Psi_{SPT} \rangle \langle \Psi_{SPT} | )
 \end{align}
This ultimately corresponds to the type-II strange correlator proposed in Ref.~\cite{LeeYouXu2022}. Our mapping indicates that the type-II strange correlator in a mixed SPT state can be traced back to the averaged strange correlation function within the purified state in the extended Hilbert space. We anticipate that our perspective offers an alternative avenue for addressing the characterization of mixed-state SPT phases. A thorough examination of this topic will be presented in our forthcoming work.

\section{summary and future directions}

We address the characterization of exotic mixed states from the lens of purification, with a summary of our main results displayed in Fig.~\ref{all}.  In particular, we have shown that a mixed state ensemble, exhibiting a strong-to-weak symmetry breaking transition, admits a purification showcasing symmetry protected topological (SPT) order in the limit of strong decoherence (error rate $p=1/2$). This perspective allows us to relate the SSSB transition recently introduced in the literature and characterized by non-vanishing Rényi-2 correlators as well as type-II strange correlators and fidelity measures; to a non-vanishing measurement-induced mutual information ---intrinsic to SPT phases. As we prove, the latter is lower-bounded by the average strange correlators of the purified state. Moreover, as we mentioned in the main text our findings reveal the interrelationship between the average strange correlator and fidelity, yet the specific nature of this averaging process remains an open question. 

In the first sections, we focus on purification states with discrete strong $\mathbb{Z}_2^S$ and weak $\mathbb{Z}_2^A$ symmetries considering three different scenarios where: (1) Both $\mathbb{Z}_2^S$ and $\mathbb{Z}_2^A$ are global ($0$-form) symmetries, (2) $\mathbb{Z}_2^S$ is global but $\mathbb{Z}_2^A$ is a $1$-form symmetry, and (3) $\mathbb{Z}_2^S$ is $1$-form while $\mathbb{Z}_2^A$ remains a global symmetry. Further employing this perspective, we also discuss the purification aspects of mixed $\mathbb{Z}_2$ topological order under the condensation of both $e$ and $f$ particles in the error-doubled field. This field has recently been shown to lead to intrinsic mixed-state topological order \cite{wang2023intrinsic}. In these cases, the purified state is defined on a 3D manifold with an open boundary, where the mixed-state density matrix is akin to the reduced density matrix of the top surface layer. The strong 1-form symmetry corresponds to the deconfined charge excitation on the surface, while the weak 1-form symmetry originates from flux loop excitations in the bulk that terminate at the top surface as a point defect. This approach directly connects to and extends the results of Ref.~\cite{sohal2024noisy}, where a similarity was noted with boundary topological order in $(3+1)$D.
Finally, we briefly extend the discussion of SSSB to a general onsite symmetry group. We also explore the landscape of $U(1)$ strong SSSB with quasi-long-range order in 2D by purifying it into a quantum spin Hall-likestate. However, a detailed construction and analysis of the corresponding $U(1)$ symmetric quantum channel is reserved for future work.
 
We conclude by illustrating several promising future research directions. We explored the spontaneous phase transition from strong to weak symmetries triggered by local quantum channels, and found that this can relate to thermal phase transitions of underlying stat-mech models (although such exact mapping is not expected to hold in general). From this perspective, we found that no finite error rate $p<1/2$ can lead to strong-to-weak SSB in $1$D (as long as the channels are local). In general, we also learned that such transitions ---corresponding to finite critical error rates $p_c$--- can only occur if the number of constraints is sufficiently large, as it e.g., happened in a 2D system with global $\mathbb{Z}_2$ symmetry, but not in the presence of subsystem symmetries. However, a finite threshold is expected in 3D\cite{xu2004strong}. Moreover, from our discussion of continuous $U(1)$ symmetries, a pertinent question is whether a Hohenberg-Mermin-Wagner-like theorem exists that bounds the occurrence of symmetry breaking with the dimensionality of the many-body system. Similarly, discovering new types of criticality triggered by quantum channels in open systems and the characterization of such critical points (from a condensed matter perspective) constitutes a promising direction.

As we discussed in Sec.~\ref{sec:econ}, the mixed-state topological order transition triggered by decoherence in the noisy toric code (studied in Refs.\cite{bao2023mixed,fan2023diagnostics,lee2022symmetry}) can be mapped to the strong-to-weak SSSB considered in Sec.\ref{sec:2d} through F-S duality. Here, the Rényi-$2$ correlator relates to the anyon condensation order parameter in terms of the vectorized density matrix. Understanding the correspondence of various other information-theoretical quantities, such as coherent information and negativity defined for mixed states, remains an intriguing open question~\cite{chen2024unconventional}. 

We are aware of another work~\cite{fidpaper} that explores the transition from strong-to-weak symmetry breaking, which would appear on arXiv on the same day.

\acknowledgements 
We are grateful to Jong-Yeon Lee, Frank Pollmann, Andrew Potter, Shengqi Sang, Romain Vasseur, Ruben Verresen, and Yifan Zhang for helpful discussions and feedback. We especially thank Jianhao Zhang and Chong Wang for explaining the relation between optimal purification and fidelity.
This work was performed in part at Aspen Center for Physics (PS, MO, YY), which is supported by National Science Foundation grant PHY-2210452 and Durand Fund. This research was also supported in part by grants NSF PHY-1748958 and PHY-2309135 to the Kavli Institute for Theoretical Physics (KITP). P.S. acknowledges support from the Caltech Institute for Quantum Information and Matter, an NSF Physics Frontiers Center (NSF Grant PHY-1733907), and the Walter Burke Institute for Theoretical Physics at Caltech.
S.G. acknowledges support from an Institute for Robust Quantum Simulation (RQS) seed grant.
M. O. was partially supported by JSPS KAKENHI Grant No. JP24H00946.

\appendix

\section{Derivation of the Equivalence Between the Rényi-2 correlator and EPR-Induced Long-Range Order}
\label{app:pp}

To provide a physical interpretation of the Rényi-2 correlator, we duplicate our Hilbert space by creating two identical copies of the SPT state, denoted \(|\Psi_{\text{SPT}}\rangle^{1}\) and \(|\Psi_{\text{SPT}}\rangle^{2}\). Next, we perform a Schmidt decomposition between the ancilla and the system for each copy of the SPT wave function as:
\begin{align}
|\Psi_{SPT}\rangle^{1}= \sum_{\alpha} \lambda_{\alpha}|\alpha \rangle^1_s|\tilde \alpha \rangle^1_a, ~|\Psi_{SPT}\rangle^{2}= \sum_{\alpha} \lambda^*_{\alpha}|\alpha^* \rangle^2_s|\tilde \alpha^* \rangle^2_a
\end{align}
Here, we take the complex conjugate of the second copy of the wave function, denoted by \(|\Psi_{\text{SPT}}\rangle^{2}\). \(|\alpha \rangle_s\) denotes the Schmidt basis for the system, while \(|\tilde{\alpha} \rangle_a\) refers to the Schmidt basis for the ancillas. \(|\alpha^* \rangle\) refers to the complex conjugate of the vector.

We now project the i-th ancillae from both the first and second copies, which are positioned at the same link $(i,i+1)$, forcing their alignment in the \(S_z\) direction by projecting onto a symmetric EPR pair:
\begin{align}
   \hat{P}_{i} = \frac{1}{2}\left(|\tilde{\uparrow}^1 \tilde{\uparrow}^2\rangle +|\tilde{\downarrow}^1 \tilde{\downarrow}^2 \rangle\right)
    \left(\langle \tilde{\uparrow}^1 \tilde{\uparrow}^2| +\langle\tilde{\downarrow}^1 \tilde{\downarrow}^2|\right)
\end{align}
for every ancilla pair at link $i$. To derive the wave function after projection, we choose an alternative basis for the purified state,
\begin{align}
|\Psi_{SPT}\rangle= \sum_{\alpha,\{ \tilde{z}_i\}} q_{\alpha,\{ \tilde{z}_i\}}|\alpha \rangle_s| \{ \tilde{z}_i\} \rangle_a, 
\end{align}
The system qubits is still written in the Schmidt basis while the ancilla qubits is written in the $S_z$ basis $\{\tilde{z}_i\}$. As the density matrix of the system is $\hrho= \sum_{\alpha} |\lambda_{\alpha}|^2|\alpha \rangle \langle \alpha |$, we get: 
\begin{align}
\sum_{\{ \tilde{z}_i\}}  q_{\alpha,\{ \tilde{z}_i\}} 
q^*_{\beta,\{ \tilde{z}_i\}} =\delta_{\alpha,\beta} |\lambda_{\alpha}|^2
\end{align}

The normalized wavefunction, after this projection, is given by:
\begin{widetext}
  \begin{align}
&|\Psi \rangle_{pp} \sim \prod_i \hat{P}_{i} [(\sum_{\alpha,\{ \tilde{z}_i\}} q_{\alpha,\{ \tilde{z}_i\}}|\alpha \rangle^1_s| \{ \tilde{z}_i\} \rangle^1_a) (\sum_{\beta,\{ \tilde{z'}_i\}} q^*_{\beta,\{ \tilde{z}_i\}}|\beta \rangle^2_s| \{ \tilde{z}_i\} \rangle^2_a)]~~\rightarrow |\Psi \rangle_{pp}=\frac{1}{\sqrt{\sum_{\alpha} |\lambda_{\alpha}|^4}}\sum_{\alpha} |\lambda_{\alpha}|^2~|\alpha \rangle^1_s |\alpha^* \rangle^2_s 
\end{align}  
\end{widetext}

In the last step, we omit the ancilla degree of freedom as they form a tensor product of EPR pairs between two copies and are decoupled to the system qubits.

The projection \(\hat{P}_{i,i+1}\) on the two ancilla copies results in the charge string \((\prod^{n}_{a=0} \tilde{Z}^1_{i+a} \tilde{Z}^2_{i+a})\) being uniformly even throughout the post-projected wave function. Consequently, the post-projection state \(|\Psi \rangle_{pp}\) exhibits long-range order in the four-point correlation function:
\begin{align}
\langle \Psi |_{pp}
~Z^1_i Z^2_i  Z^1_{i+n+1}Z^2_{i+n+1}|\Psi \rangle_{pp}=1
\end{align}
We will now establish that the Rényi-\(2\) correlator precisely corresponds to the four-point correlation function within the post-projection state \(|\Psi \rangle_{pp}\). 

After tracing out the ancilla from the purified state \(|\Psi_{\text{SPT}}\rangle = \sum_{\alpha} \lambda_{\alpha}|\alpha \rangle_s|\tilde{\alpha} \rangle_a\), we obtain the density matrix \( \hrho \):
\begin{align}
&\hrho= \sum_{\alpha} |\lambda_{\alpha}|^2|\alpha \rangle \langle \alpha |
\end{align}
Tracing out the ancilla effectively involves projecting the ancilla in both the ket and bra spaces to be identical, essentially aligning them in the same $S^z$ direction.
By considering the bra vector as a duplicate copy, the ancilla tracing procedure exactly corresponds to the projection operation in Eq.~\ref{postp}. Likewise, applying the Choi-Jamiołkowski isomorphism maps the density matrix into a pure state within the doubled Hilbert space:
\begin{align}
&\hrho= \sum_{\alpha} |\lambda_{\alpha}|^2|\alpha \rangle \langle \alpha |~ \rightarrow ~|\hrho \rangle \rangle \sim \sum_{\alpha} |\lambda_{\alpha}|^2|\alpha \rangle^1 |\alpha^* \rangle^2
\end{align}
Indeed, \(|\hrho \rangle \rangle\) agrees with the post-projection wave function in Eq.~\ref{postp}. This similarity suggests that tracing out the ancilla and acquiring the density matrix mirrors the process of having a duplicated copy (originating from the ket and bra vectors) of the system, with the ancilla in both copies being projected to an EPR pair.

Drawing from this analogy, when calculating the Rényi-\(2\) correlator in Eq.~\ref{eq:swSSB}, the operators acting on the left and right sides of the density matrix \(\hrho\) can be interpreted as measuring operators for both identical copies post-ancilla projection. Consequently, we can express the Rényi-\(2\) correlator in the following way:
 \begin{align}
    &\frac{\Tr \left(Z_0 Z_i\hrho Z_0 Z_i\hrho \right)}{\Tr \hrho^2} \nonumber\\
    &= \frac{(\sum_{\alpha,\alpha'} |\lambda_{\alpha}|^2 |\lambda_{\alpha'}|^2)\langle \alpha|Z_0 Z_i|\alpha' \rangle \langle \alpha'|Z_0 Z_i|\alpha \rangle}{\sum_{\alpha} |\lambda_{\alpha}|^4} \nonumber\\
    &= \frac{(\sum_{\alpha,\alpha'} |\lambda_{\alpha}|^2 |\lambda_{\alpha'}|^2)\langle \alpha|Z_0 Z_i|\alpha' \rangle (\langle \alpha|Z_0 Z_i|\alpha' \rangle)^*}{\sum_{\alpha} |\lambda_{\alpha}|^4} \nonumber\\
    &= \frac{(\sum_{\alpha,\alpha'} |\lambda_{\alpha}|^2 |\lambda_{\alpha'}|^2)\langle \alpha|^1 \langle \alpha^*|^2 Z^1_0 Z^1_i Z^2_0 Z^2_i|\alpha' \rangle^1 |{\alpha^*}' \rangle^2}{\sum_{\alpha} |\lambda_{\alpha}|^4} \nonumber\\
    &=\langle \Psi |_{pp}
Z^1_i Z^2_i  Z^1_{i+n+1}Z^2_{i+n+1}|\Psi \rangle_{pp}
\end{align}
This precisely corresponds to the four-point correlation of the post-projected wave function.

\subsection{Derivation of the Equivalence Between the Rényi-2 correlator and spin-correlation in classical Ising model}\label{app:ising}

After applying the unitary described in Eq.~\ref{unitary2} and tracing out the ancilla, we obtain a \(\mathbb{Z}_2\) symmetric quantum channel:
\begin{align}
    & \hrho^D = \mathcal{E}[\hrho_0], \ \ \mathcal{E} = \prod_{\vect{i}} \cE_{\vect{i}}, \nonumber \\ &  \cE_{\vect{i}}[\hrho_0] =(1-p) \hrho_0 + p Z_i Z_{i+1} \hrho_0 Z_i Z_{i+1}, 
\end{align}
with an error rate $p=\frac{1-\sin(2\theta)}{2}$. Initializing in a pure product state $\hrho=\ket{\rightarrow}\bra{\rightarrow}^{\otimes N}$, we employ the Choi-Jamiolkowski isomorphism to write the resulting decohered density matrix in the vectorized form as
\begin{equation}
\begin{aligned}
    \hrho\rightarrow |\hrho\rangle \rangle &= \prod_{\langle i,j\rangle}\mathcal{E}_{i,j} |\hrho_0 \rangle \rangle 
    \\ &= \prod_{\langle i,j\rangle} \left[(1-p) + p Z_{i}Z_{j}\otimes Z_{i}  Z_{j}\right]\ket{\rightarrow}^{\otimes N}\ket{\rightarrow}^{\otimes N}\\
    & \propto \exp \left( \beta \sum_{\langle i,j\rangle}Z_{i}Z_{j}\otimes Z_{i}  Z_{j}  \right)\ket{\rightarrow}^{\otimes N}\ket{\rightarrow}^{\otimes N}
\end{aligned}
\end{equation}
up to an overall $p$-dependent factor, and with $\tanh(\beta)=\frac{1-\sin(2\theta)}{1+\sin(2\theta)}$. One can then write $\ket{\rightarrow}=(\ket{0}+\ket{1})/\sqrt{2}$ in the local $Z$-basis such that we finally obtain
\begin{align}
    |\hrho\rangle \rangle & =2^{-N}\sum_{\sigma, \sigma^\prime} e^{ \beta \sum_{\langle i,j \rangle} \sigma_i \sigma_j  \sigma^\prime_i \sigma^\prime_j } \ket{\sigma}\ket{\sigma^\prime},
\end{align}
which can be rewritten as
\begin{align}
     |\hrho\rangle \rangle & =2^{-N}\sum_{\sigma, \sigma^\prime} e^{ \beta \sum_{\langle i,j \rangle} \sigma_i \sigma_j   } \ket{\{\sigma_i\sigma_i^\prime\}} \ket{\sigma^\prime}.
\end{align}
Hence, diagonal expectation values on the local $Z$ basis are given by thermal expectation values evaluated on the classical Ising model. In particular, one finds that the Rényi-2 correlator maps to the 2-point thermal correlation of the Ising model at temperature $2\beta$, i.e.,
\begin{equation} \label{eq:Renyi2_to_Ising}
    \frac{{\rm tr}\left(Z_a Z_b\hrho Z_a Z_b\hrho \right)}{{\rm tr}\hrho^2}=\frac{1}{Z}\sum_{\sigma} e^{ 2\beta \sum_{\langle i,j \rangle} \sigma_i \sigma_j }\sigma_a \sigma_b
\end{equation}
 This implies that in one dimension, the decohered density matrix lacks long-range order in Rényi-2 correlator for any \(p < \frac{1}{2}\). In two dimensions, it triggers a phase transition at finite \(\beta\) that is akin to the 2D classical Ising transition at finite temperature.

\section{Derivation for averaged strange correlator}\label{app:sc}

We begin by examining the strange correlator of the 2D purified wave function in Eq.~\ref{unitary2dweak},
\begin{align}
   &  O^s(x,y)=\frac{\langle \Psi_{trivial} | Z(x) Z(y) | \Psi_{SPT}(\theta) \rangle}{\langle \Psi_{trivial} | \Psi_{SPT}(\theta)\rangle}\nonumber\\
   &\ket{\Psi_{SPT}(\theta)}=\sum_{\{ z_i \}} \prod_{\langle i,j\rangle} \left( \frac{1+z_iz_j}{2}(\cos{\theta}|s_{ij}=1\rangle -\sin{\theta}|s_{ij}=-1\rangle )\right.
   \nonumber\\
   &\left.+\frac{1-z_iz_j}{2}(\sin{\theta}|s_{ij}=1\rangle -\cos{\theta}|s_{ij}=-1\rangle ) \right)  |\{ z_i \}\rangle \nonumber\\
      & \ket{\Psi_{trivial}(s_{ij})}=~  \otimes_{\langle i,j \rangle}|s_{ij}=1  \rangle \otimes_i | x_i=1 \rangle 
\end{align}
In this context, \( s_{ij} \) represents ancilla qubits (in the \( Z \) basis), situated on the links of the square lattice between the nearest vertices at sites \( i \) and \( j \). The term \( z_i (x_i) \) denotes the spin pattern of the system qubits at vertex $i$ in the \( Z(X) \) basis. $|\{ z_i \}\rangle$ refers to a specific many-body pattern and we sum over all possible patterns.
For \( \theta = 0 \), \( \ket{\Psi_{SPT}(0)} \) corresponds to the SPT state, which exhibits exact 1-form \( Z_2^A \) symmetry (acting on the ancilla) and 0-form \( Z_2^B \) symmetry (acting on the system), as defined in Eq.~\ref{sym2d}. When \( \theta \) is nonzero, the ancilla spin on the link tilts, leading to the absence of the exact 1-form \( Z_2^A \) symmetry. At \( \theta = \pi/4 \), the ancilla spins decouple from the system qubits. The trivial wave function \( \Psi_{trivial} \) is symmetrically chosen, with the ancilla polarized in the \( S^z \) direction as \( |s_{ij} = 1\rangle \) and the system qubits polarized in the \( S^x \) direction as \( |x_i = 1\rangle \).
We can rewrite the denominator of the strange correlator as:
\begin{align}
   &  \langle \Psi_{trivial} | \Psi_{SPT}(\theta) \rangle=   \langle  x_a=1  | \hat{P}(s_{ij}=1)\Psi_{SPT}(\theta) \rangle
\end{align}
The projection $| \hat{P}(s_{ij}=1)\Psi_{SPT}(\theta) \rangle$ projects each ancilla qubit in $\Psi_{SPT}(\theta)$ into the $s_{ij}=1$ state. The post-measurement wave function for the system qubits then takes the form:
\begin{align}
   &  | \hat{P}(s_{ij}=1)\Psi_{SPT}(\theta) \rangle \nonumber\\
   &=\sum_{\{ z_i \}}  \prod_{\langle i,j\rangle}( \frac{1+z_iz_j}{2} \cos{\theta}+\frac{1-z_iz_j}{2}\sin{\theta} )|\{ z_i \}\rangle \nonumber\\
     &=\sum_{\{ z_i \}}
     \frac{e^{\tilde{\beta}\sum_{\langle i,j\rangle}z_i z_j}}{(e^{-2 \tilde{\beta}}+e^{2 \tilde{\beta}})^{N/2}}|\{ z_i \}\rangle
\end{align}
With $e^{-2\tilde{\beta}} = \tan(\theta)$ (note this $\beta,\theta$ relation is different from the Rényi-2 correlator discussed in Sec.~\ref{sec:2d}), and with $i, j$ only running over nearest neighbor sites. This maps the projected wave function to the partition function of the Ising model. Likewise, the product state $\prod_a \otimes | x_a=1 \rangle$ can be written as follows:
\begin{align}
   & \otimes_i  | x_i=1 \rangle =\sum_{\{ z_i \}}
     \frac{1}{(2)^{N/2}}|\{ z_i \}\rangle
\end{align}
This is akin to the Ising model at infinite temperature. Based on this notation, we can rewrite the denominator of the strange correlator as follows:
\begin{align}
    \langle \Psi_{trivial} | \Psi_{SPT}(\theta) \rangle =
     \frac{\sum_{\{ z_i \}} e^{\tilde{\beta}\sum_{\langle i,j\rangle}z_i z_j}}{(2e^{-2\tilde{\beta}}+2e^{2 \tilde{\beta}})^{N/2}}
\end{align}
This is akin to the partition function of the Ising model. Likewise, the numerator can be expressed as follows:
\begin{align}
   &  \langle \Psi_{trivial} |Z(x) Z(y)| \Psi_{SPT}(\theta) \rangle= \nonumber\\
    &=\sum_{\{ z_i \}} \frac{z_x z_y}{(2)^{N/2}}\prod_{\langle i,j \rangle}( \frac{1+z_iz_j}{2} \cos{\theta}+\frac{1-z_iz_j}{2}\sin{\theta} ) \nonumber\\
     &=\sum_{\{ z_i \}}
     \frac{z_x z_ye^{\tilde{\beta}\sum_{\langle i,j\rangle}z_i z_j}}{(2e^{-2 \tilde{\beta}}+2e^{2 \tilde{\beta}})^{N/2}}.
\end{align}
Based on this formulation, the strange correlator for $\ket{\Psi_{SPT}(\theta)}$ corresponds to the spin-correlator of the 2D Ising model at an effective temperature $-2\tilde{\beta} = \ln(\tan(\theta))$.

Now we consider measuring the \textit{disorder-averaged strange correlator} by choosing various trivial states with a random assortment of ancilla spins \( \{ s_{ij}=\pm 1\} \), and then averaging over these random selections.
\begin{align}
    &\bar{O}^s(x,y)=\sum_{\{ s_{ij} \}}\frac{\langle \Psi_{trivial}(\{ s_{ij} \}) | Z(x) Z(y) | \Psi_{SPT} \rangle}{\langle \Psi_{trivial}(\{ s_{ij} \}) | \Psi_{SPT} \rangle} \nonumber\\
      & \Psi_{trivial}(\{ s_{ij} \})=~  \otimes_{\langle ij \rangle}|s_{ij}  \rangle \otimes_i | x_a=1 \rangle 
 \end{align}
 The trivial state $\ket{\Psi_{trivial}(\{ s_{ij} \})}$ is a direct product of system spins polarized in the $S^x$ direction and a random assortment of ancilla spins \( \{ s_{ij}=\pm 1\} \) polarized in the $\pm S^z$ directions. Here, we average over all \( \{ s_{ij}=\pm 1\} \) with the same probability. Such an averaged strange correlator is akin to the two-point correlator of the random bond Ising model, which vanishes at large distances $\bar{O}^s(x,y)=0$.

\subsection{Annealed disorder average strange correlator and type-II strange correlator}

We introduce the \textit{annealed disorder average} for the strange correlator: 
\begin{align}
   & \langle \overline{O^s(x,y)} \rangle ^2=\frac{\sum_{\{ s_{ij} \}}|\langle \Psi_{trivial}(\{ s_{ij} \}) | Z(x) Z(y) | \Psi_{SPT} \rangle|^2}{\sum_{\{ s_{ij} \}}|\langle \Psi_{trivial}(\{ s_{ij} \}) | \Psi_{SPT} \rangle|^2}\nonumber\\
 &= \frac{\sum_{\{ s_{ij} \}}\sum_{\{ z^1_i, z^2_i \}}(z^1_x z^1_y z^2_x z^2_y)e^{\tilde{\beta} \sum_{\langle i,j\rangle}s_{ij}(z^1_i z^1_j+z^2_i z^2_j)}}{\sum_{\{ s_{ij} \}} \sum_{\{ z^1_i, z^2_i \}}e^{\tilde{\beta} \sum_{\langle i,j\rangle}s_{ij}(z^1_i z^1_j+z^2_i z^2_j)}}
\end{align}
The annealed average for the strange correlator can be treated as two copies of Ising models, with a measurement of the spin correlator acting on both copies. We define a gauge transformation $\tau_i = z^1_i z^2_i, s'_{ij} = s_{ij} z^2_i z^2_j$, and consequently, the annealed average in becomes:
\begin{align}\label{appeq:annealeising}
   & \langle \overline{O^s(x,y)} \rangle ^2= \frac{\sum_{\{ s'_{ij} \}} \sum_{\{ \tau_{i} \}}e^{\tilde{\beta} \sum_{\langle i,j\rangle}s'_{ij}}(\tau_x \tau_y)e^{\tilde{\beta} \sum_{\langle i,j\rangle}s'_{ij}\tau_i \tau_j}} {\sum_{\{ s'_{ij} \}} \sum_{\{ \tau_{i} \}}e^{\tilde{\beta}\sum_{\langle i,j\rangle}s'_{ij}}e^{\tilde{\beta} \sum_{\langle i,j\rangle}s'_{ij}\tau_i \tau_j}}
\end{align}
Eq.~\ref{appeq:annealeising} is akin to the \textit{annealed average} of the two-point correlation function in the random bond Ising model (RBIM) along the Nishimori Line\cite{nishimori1981internal}, 
where the disordered bond coupling $s'_{ij}$ has the probability distribution $\to P(s'_{ij}=1) = \frac{e^{\tilde{\beta}}}{e^{-\tilde{\beta}} + e^{\tilde{\beta}}}$.
One can then easily show that the annealed average of the random bond Ising model along Nishimori Line can be mapped to the 2D classical Ising model. In particular, the partition function appearing in the denominator of Eq.~\ref{appeq:annealeising} can be exactly expressed as
\begin{equation}
    \begin{aligned}
        & \sum_{\{ \tau_{i} \}}\prod_{\langle i,j\rangle} \left(\sum_{ s'_{ij} =\pm 1} e^{\tilde{\beta}s'_{ij}}e^{\tilde{\beta} s'_{ij}\tau_i \tau_j}\right)\\
        &=\left(\frac{\cosh^2(\tilde{\beta})}{\cosh(\beta)}\right)^{N_{\textrm{links}}}\sum_{\{\tau_i\}}e^{\beta \sum_{\langle i,j\rangle} \tau_i\tau_j},
    \end{aligned}
\end{equation}
with $\beta$ coinciding with that of the Rényi-II correlator and relates to $\tilde{\beta}$ via $\tanh(\beta)=\tanh^2(\tilde{\beta})$. Hence, $\langle \overline{O^s(x,y)} \rangle ^2$ corresponds to the thermal correlation function evaluated on the $2D$ Ising model on the square lattice at inverse temperature $\beta$. In the following, we will recover this result in an alternative manner.

Notably, the annealed averaged strange correlator of the purified state \(\Psi_{SPT}\) can be mapped to the type-II strange correlator of the mixed state after the ancilla has been traced out. While the type-II strange correlator was originally proposed for decoherent SPT mixed states in Ref.~\cite{LeeYouXu2022}, our findings reveal that it stems from the annealed-average strange correlation in the enlarged Hilbert space resulting from purification. Given that the operators \(Z(x) Z(y)\) in the strange correlator act only on the system's qubits, we can alternatively express Eq.~\ref{annealed} as:
\begin{align}\label{type2}
    &\langle \overline{O^s(x,y)} \rangle ^2=\frac{\sum_{\{ s_{ij} \}} |\langle \Psi_{trivial}(\{ s_{ij} \}) | Z(x) Z(y) | \Psi_{SPT} \rangle|^2}{\sum_{\{ s_{ij} \}} |\langle \Psi_{trivial}(\{ s_{ij} \}) | \Psi_{SPT} \rangle|^2}\nonumber\\
     &= \frac{\sum_{\{ s_{ij} \}} |\langle s_{ij}| \otimes  \langle x_i=1| Z(x) Z(y) | \Psi_{SPT} \rangle|^2}{\sum_{\{ s_{ij} \}} |\langle s_{ij}| \otimes  \langle x_i=1| \Psi_{SPT}\rangle|^2}\nonumber\\
       &= \frac{\sum_{\{ s_{ij} \}}  |\langle x_i=1| Z(x) Z(y) \langle s_{ij}| \Psi_{SPT} \rangle|^2}{\sum_{\{ s_{ij} \}} \langle x_i=1 |\langle s_{ij}|\Psi_{SPT}\rangle|^2}\nonumber\\
       &=\frac{\Tr[\hrho^{0} Z(x) Z(y) \hrho Z(x) Z(y)] }{\Tr[\hrho^{0}  \hrho ]}\nonumber\\
 \end{align}
 where
 \begin{equation}
 \begin{aligned}
     \hrho&=\Tr_{\text{ancilla}}( |\Psi_{SPT} \rangle \langle \Psi_{SPT} | ) \propto \sum_{\sigma,\sigma^\prime} e^{\beta \sum_{\langle i,j\rangle} \sigma_i \sigma_j \sigma^\prime_i \sigma^\prime_j}\ket{\sigma}\bra{\sigma^\prime}
 \end{aligned}
 \end{equation}
 with $\tanh(\beta)=(1-\sin(2\theta))/(1+\sin(2\theta))$ (Same as for Rényi-II correlator), and $\hrho^{0}= \otimes_i|x_i=1\rangle \langle x_i=1 |$. 
The annealed average of the strange correlator in the purified state $\Psi_{SPT}$ exactly maps to the type-II strange correlator for the mixed state $\hrho$ (and agrees with the Rényi-II correlator although for double temperature). Here, $\hrho^{0}$ is chosen to be a pure state density matrix consisting of a trivial tensor product state $|x_i=1\rangle$, and $\hrho$ refers to the density matrix of the system obtained after tracing out the ancilla from $\Psi_{SPT}$. Eq.~\ref{type2} can be mapped to the thermal correlation of Ising model:
\begin{align}
    &\langle \overline{O^s(x,y)} \rangle ^2=\frac{1}{\mathcal{Z}(\beta)}\sum_{\{\sigma\}} \sigma_x \sigma_y e^{\beta \sum_{\langle i,j \rangle} \sigma_i \sigma_j}.
 \end{align}

\section{MIE upper-bounds for annealed averaged strange correlators}\label{app:mi}

The mutual information between two distant qubits, \(A\) and \(B\), provides an upper bound for connected correlators between operators supported in these regions. In this appendix, we demonstrate that the measurement-induced mutual information, as proposed in Ref.~\cite{cheng2023universal}, also imposes upper bounds on \emph{strange} correlators. Following the proof and setup in Ref.~\cite{cheng2023universal}, we consider a system of qubits \(|\psi\rangle\) that is partitioned into three regions: \(A\), \(B\), and \(C\), along with the ancilla qubits \(s_{ij}\), as follows:
\begin{align}
	|\Psi_{SPT}\rangle = \sum_{a,b,c, s_{ij}}\phi_{abc,s_{ij}}|abc,s_{ij} \rangle,
\end{align}
Here, \(a\) and \(b\) represent the two distant qubits on which we will apply operators to examine the strange correlator, while \(c\) labels states in the remaining degrees of freedom. We perform measurements on region \(C\) in conjunction with the ancilla \(\{s_{ij}\}\), yielding the outcome \(|m_c, s_{ij} \rangle\). The resulting wavefunction is then given by:
\begin{align}
	|\psi_{c,s_{ij}}\rangle=\frac{\langle m_c, s_{ij} |\psi\rangle}{\sqrt{p_{m_c,s_{ij}}}}=\sum_{a,b}\frac{\phi_{ab,m_c,s_{ij}}}{\sqrt{p_{m_c,s_{ij}}}}\vert ab\rangle,
\end{align}
with probability $p_{m_c,s_{ij}}=\sum_{a,b}\vert\phi_{ab,m_c,s_{ij}}\vert^2$. 

In the same context, we can also define a strange correlator between a trivial tensor product state \(\vert m\rangle=\vert m_am_bm_c, s_{ij}\rangle\) and the purified state \(\vert\Psi_{SPT}\rangle\):
\begin{align}
	C(m_a,m_b,m_c,s_{ij})=\frac{\langle m| O_A O_B\vert\psi\rangle}{\langle m|\psi\rangle}
\end{align}
Here, \(m_a\) and \(m_b\) denote states in regions \(A\) and \(B\) that carry a fixed charge, while \(O_A\) and \(O_B\) are charged local operators in regions \(A\) and \(B\), respectively.

To demonstrate the relationship between strange correlators and measurement-induced mutual information, we initially represent the mutual information in terms of relative entropy:
\begin{align}
	&I(A,B)[\vert\psi_{c,s_{ij}}\rangle]=S(\hrho^{c,s_{ij}}_{AB}\vert \hrho^{c,s_{ij}}_A\otimes\hrho^{c,s_{ij}}_B),\nonumber\\
&\hrho_{AB}^{c,s_{ij}}=\vert\psi_{c,s_{ij}}\rangle\langle\psi_{c,s_{ij}}\vert,\nonumber\\
&\hrho_{A/B}^{c,s_{ij}}=\tr_{B/A}(\vert\psi_{c,s_{ij}}\rangle\langle\psi_{c,s_{ij}}\vert)
\end{align}
Using the norm bound $S(\hrho\vert\sigma)\geq\frac{1}{2}\vert\vert\hrho-\sigma\vert\vert^2_1$ and the trace inequality $\vert\vert X\vert\vert_1\geq\tr(XY)/\vert\vert Y\vert\vert$, we obtain: 
\begin{align}
	I(A,B)[\vert\psi_{c,s_{ij}}\rangle]\geq\frac{1}{2}\frac{(\tr(\hrho_{AB}^{c,s_{ij}} Y)-\tr(\hrho_A^{c,s_{ij}}\otimes\hrho_B^{c,s_{ij}} Y))^2}{\vert\vert Y\vert\vert^2}.
\end{align}
Let $Y=\vert m_am_b\rangle\langle m_am_b\vert O_A O_B$, then the right-hand side of the inequality becomes
\begin{align}~\label{aeq:SC}
	&I(A,B)[\vert\psi_{c,s_{ij}}\rangle]\geq \frac{p_{m_am_bm_c,s_{ij}}^2}{2\overline{O}^2_A\overline{O}^2_B p_{m_c,s_{ij}}^2}|C(m_a,m_b,m_c,s_{ij})|^2\nonumber\\
 &>\frac{p_{m_a m_b}^2}{2\overline{O}^2_A\overline{O}^2_B}|C(m_a,m_b,m_c,s_{ij})|^2
 \end{align}
where \(\overline{O}_{A/B}\) represents the norm of the operator. This suggests that the strange correlator of the post-measurement state \(\vert\psi_{c,s_{ij}}\rangle\) provides a lower bound for the measurement-induced mutual information. If we sum over different ancilla patterns,
\begin{align}
	&\frac{\sum_{s_{ij}}  p_{m_a,m_b,m_c,s_{ij}} I(A,B)[\vert\psi_{c,s_{ij}}\rangle]}{  p_{m_a,m_b,m_c}}\nonumber\\
 &> c_0 \frac{(\sum_{s_{ij}}  p_{m_a,m_b,m_c,s_{ij}}|C(m_a,m_b,m_c,s_{ij})|^2)}{p_{m_a,m_b,m_c}}\nonumber\\
 &=c_0\langle \overline{O^s(x,y)} \rangle ^2
 \end{align}
where \(c_0\) is a non-universal constant that depends on the choice of \(p_{m_a m_b}\) and \(O_{A/B}\). The first line can be interpreted as the disorder-averaged measurement-induced mutual information by projecting the purified state onto different \(s_{ij}\) patterns and averaging the mutual information across these projections. Thus, the measurement-induced mutual information is lower-bounded by the annealed average of the strange correlator.

\section{SSSB of general on-site symmetries}
\label{app:general}


Finally, we discuss the mixed state strong-to-weak symmetry breaking for more
general global on-site symmetries associated with a finite symmetry group $G$.
A local Hilbert space of dimension $d$ is associated with each site, forming a
$d$-dimensional unitary representation $u_g$, where $g\in G$.
The order (the number of elements) of $G$ is denoted as $|G|$.
$u_g$ is not necessarily irreducible, but can be decomposed into a direct sum of irreducible representations.
Let us denote the representation of $g$ acting on site $i$ as $u_g^{(i)}$.
The global transformation is then
given by $U_g = \prod_i u_g^{(i)}$, where $i$ labels the lattice sites in
the total system $\Lambda$.
The size of $\Lambda$ (number of sites) in $\Lambda$ is denoted as $|\Lambda|$.

For the identity element $e \in G$,
$u_e = \mathbb{I}_d$.
There may be other elements $g \in G$ which has the representation of the form
$u_g = e^{i \theta_g}\mathbb{I}_d$.
The set of all such elements $g$ form a subgroup $H$ of $G$.
For any given element $h \in H$, 
\begin{align}
u_{g^{-1}h g} = u_{g^{-1}} u_h u_g = {u_g}^\dagger e^{i\theta_h} \mathbb{I}_d u_g = e^{i\theta_h} \mathbb{I}_d
\end{align}
and thus $g^{-1} h g \in H$ for any $g \in G$.
This means that $H$ is the normal subgroup of $G$, and we can
define the quotient group $\tilde{G} \equiv G/H$.
By construction, the identity $e$ is the only element of $\tilde{G}$ which
has the representation of the form $u_g = e^{i \theta_g} \mathbb{I}_d$.

In a unitary representation $u_g$,
all the eigenvalues $\{ \lambda_a \}$ $(a=1,2,\ldots,d)$ of $u_g$
have absolute value unity $|\lambda_a|=1$.
We can define $d$-dimensional complex vectors $\vec{\lambda}=(\lambda_1,\lambda_2,\ldots,\lambda_d$, and $\vec{I}=(1,1,\ldots,1)$.
We have
\begin{align}
\tr{u_g} = & {}^t\vec{I}^*\cdot \vec{\lambda}
\end{align}
Cauchy-Schwarz inequality implies
\begin{align}
 \left| {}^t\vec{I}^*\cdot \vec{\lambda} \right|^2 \leq &
    \left| \vec{I} \right| \left|\vec{\lambda}\right| = d^2
\end{align}
Therefore
\begin{align}
    \left| \tr{u_g}\right| \leq d ,
\end{align}
and the equality holds if and only if $\vec{\lambda} \propto \vec{I}$,
which implies
\begin{align}
    \lambda_a =& e^{i\theta} ,
\end{align}
for $\theta$ independent of the index $a$.
This is equivalent to
\begin{align}
    u_g = & e^{i \theta} \mathbb{I}_d .
\end{align}

Therefore, for the quotient group $\tilde{G}=G/H$,
\begin{equation}
     |\tr{u_g} | < d  \;\;\; \text{if $g \in \tilde{G} \neq e$} .
     \label{eq:tru_tilG}
\end{equation}
Since any $h \in H$ satisfies $h^{|H|}=e$ (by the virtue of $H$ being a finite group),
$|H| \theta_h$ is an integral multiple of $2\pi$.
For convenience let us choose the system size $|\Lambda|$ to be an integral multiple of $|H|$.
Then for $h \in H$, $U_h = \mathbb{I}_{|\Lambda|}$.

Following the construction for the case of the $\mathbb{Z}_2$ symmetry, we construct the ``spontaneous strong $G$-symmetry breaking''
mixed state as follows.
For each basis state $| s_\Lambda \rangle$ of the entire system, we consider the symmetric projection
\begin{align}
  P^0 | s_\Lambda \rangle = \frac{1}{|G|} \sum_{g \in G} U_g | s_\Lambda \rangle 
  = \frac{1}{|\tilde{G}|} \sum_{g \in \tilde{G}} U_g | s_\Lambda \rangle 
\end{align}
$P^0$ is the projection operator to the trivial representation sectors of $G$ in the total Hilbert space.
Then we can define the symmetric pure state
\begin{equation}
  |\Psi_S \rangle = C_s \sum_{s_\Lambda} P^0 | s_\Lambda \rangle ,
\end{equation}
where $C_s$ is the normalization factor so that $\langle \Psi_S | \Psi_S \rangle = 1$.
By construction, $|\Psi_S \rangle$ is invariant under the global symmetry transformation $U_g$ for all $g\in G$:
\begin{equation}
  U_g |\Psi_S \rangle = |\Psi_S \rangle .
\end{equation}
The density matrix of the SSSB mixed state,
analogous to Eq.~\eqref{densitybreak} for $G=\mathbb{Z}_2$ (i.e., at $p=1/2$)
is constructed as a projection to the pure state $|\Psi_S \rangle$:
\begin{equation}
  \hrho_S = | \Psi_S \rangle \langle \Psi_S | .
\end{equation}
By construction, $\hrho_S$ has the strong $G$-symmetry
\begin{equation}
  U_g \hrho_S = \hrho_S {U_g}^\dagger = \hrho_S .
\end{equation}
Alternatively, $\hrho_S$ can be also written as
\begin{align}
  \hrho_S = &
  |C_s|^2 \sum_{s_\Lambda} P^0 | s_\Lambda \rangle \langle s_\Lambda | P^0 
  \nonumber \\
  = & |C_s|^2 P^0 = \frac{|C_s|^2}{|\tilde{G}|} \sum_{g \in \tilde{G}} U_g .
  \label{eq:rho_S}
\end{align}
Since $\Tr{P^0}$ is the number of identity representations $N_I$ appearing in the
total Hilbert space, $C_s = 1/\sqrt{N_I}$.

Using the on-site nature of the symmetry transformation, we also see
\begin{align}
 N_I = \Tr{P^0} = & \frac{1}{|\tilde{G}|}
 \sum_{g \in \tilde{G}} \left( \tr{u_g} \right)^{|\Lambda|} .
\end{align}
In the thermodynamic limit $|\Lambda|\to \infty$, the sum is dominated by
$g = e \in \tilde{G}$ thanks to Eq.~\eqref{eq:tru_tilG}.
Then
\begin{align}
    \sum_{g \in \tilde{G}} \left( \tr{u_g} \right)^{|\Lambda|} \sim d^{|\Lambda|}
    \label{eq:tru_power}
\end{align}
in the thermodynamic limit  $|\Lambda|\to \infty$.



The normalized density matrix in the thermodynamic limit is thus 
\begin{align}
  \hrho_S \sim \frac{1}{d^{|\Lambda|}} \sum_{g \in \tilde{G}} U_g 
  = \frac{|\tilde{G}|}{d^{|\Lambda|}} P^0 .
\end{align}
This implies that the number of identity representation sectors in the total Hilbert space is
asymptotically
\begin{align}
  N_I \sim \frac{d^{|\Lambda|}}{|\tilde{G}|} ,
  \label{eq:N_id}
\end{align}
where $|\tilde{G}|=|G|/|H|$.

Now let us introduce an order parameter $\calO$.
While observables in quantum mechanics are Hermitian operators, for convenience we allow non-Hermitian operators, which are
given by a complex linear combination of Hermitian operators,
as order parameters.
The order parameter $\calO$ should be ``charged'' under the symmetry $G$.
Generally, there is a multiplet of order parameters $\calO^\alpha$
defined on a local spatial region $\Omega$ which transforms as
\begin{align}
  U_g^\dagger \calO^\alpha U_g = \sum_{\beta} M_{\alpha\beta}(g) \calO^\beta ,
\end{align}
where $M_{\alpha\beta}(g)$ is a unitary representation of
$G$ which does not include the identity representation.
Without losing generality, let us assume that $M_{\alpha\beta}$
is an irreducible representation of $G$.
There is a conjugate order parameter $\bar{\calO}_j$ which transforms by the conjugate representation $M^*_{\alpha\beta}(g)$.
Since for $h \in H$ the transformation is trivial
$U_h^\dagger \calO^\alpha U_h = \calO^\alpha$, we can replace $G$ with
the quotient group $\tilde{G}$ in the above.

For any density matrix $\hrho$ which is weakly symmetric under $G$,
the expectation value of the order parameter should vanish.
Formally, this can be shown as follows.
\begin{align}
  \langle \calO^\alpha \rangle = & \Tr{\hrho \calO^\alpha} =
  \Tr{ U_g \hrho U^\dagger_g \calO^\alpha}
  \notag \\
  = & \sum_\beta M_{\alpha\beta}(g) \langle \calO^\beta \rangle ,
\end{align}
for any $g \in G$.
Summing over each side over $g \in G$,
\begin{align}
  |G| \langle \calO^\alpha \rangle = &
  \sum_\beta \sum_{g \in G} M_{\alpha\beta}(g) \langle \calO^\beta \rangle = 0 ,
  \label{eq:charged_vanish}
\end{align}
since
\begin{align}
  \sum_{g \in G} M_{\alpha\beta}(g) = 0 ,
\end{align}
for any irreducible, non-identity representation $M_{\alpha\beta}$ of $G$
thanks to the great orthogonality relations.

As we have seen above, the vanishing of the charged order parameter only requires the weak symmetry of the density matrix
and the strong symmetry does not make a difference at this level.
However, in Rényi-2 expectation values, the strong symmetry plays an important role.
Let us consider the expectation value
\begin{align}
  \Tr{\hrho \bar{\calO}^\alpha \hrho \calO^\beta } 
\end{align}
If $\hrho$ is weakly symmetric, it follows that 
\begin{align}
  \Tr{ \hrho \bar{\calO}^\alpha \hrho \calO^\beta} = &
  \Tr{ U_g \hrho U^\dagger_g \bar{\calO}^\alpha U_g \hrho U^\dagger_g \calO^\beta }
  \notag \\
  = & \sum_{\gamma\delta} M^*_{\alpha\gamma}(g) M_{\beta\delta}(g)
  \Tr{\hrho \bar{\calO}^\gamma \hrho \calO^\delta } .
\end{align}
Thus the order parameters form a tensor product representation $M^* \otimes M$ of $G$.
Since
\begin{align}
  \sum_{g \in G} \tr{M^*(g) \otimes M(g)} = & \sum_{g \in G} \left|\tr{M(g)}\right|^2 > 0,
\end{align}
the tensor product representation must contain the identity representation.
Thus the Rényi-2 expectation value is generally non-vanishing (i.e. there is no requirement for it to vanish from the symmetry).

However, if the density matrix is strongly symmetric, we can transform one of the order parameters only: 
\begin{align}
  \Tr{ \hrho \bar{\calO}^\alpha \hrho \calO^\beta } = &
  \Tr{ \hrho \bar{\calO}^\alpha \hrho U^\dagger_g \calO^\beta U_g }
  \notag \\
  = & \sum_{\gamma} M_{\beta\gamma}(g) \Tr{\hrho \bar{\calO}^\alpha \hrho \calO^\beta } .
\end{align}
Thus the Rényi-2 expectation value vanishes, following the same logic as in Eq.~\eqref{eq:charged_vanish}.
Nevertheless, the Rényi-2 \emph{correlation function} can be non-vanishing and signal a long-range order,
as much as the conventional correlation function of order parameters do in conventional ordered phases.


In order to discuss the correlation function, we have to invoke the locality
of the order parameters:
the local order parameter $\calO^\alpha_\Omega$ is defined on a local region
$\Omega \subset \Lambda$.
The standard correlation function in the SSSB mixed state~\eqref{eq:rho_S}
between $\bar{\calO}^\alpha_{\Omega_1}$ and $\calO^\beta_{\Omega_2}$
in the SSSB mixed state~\eqref{eq:rho_S} reads
\begin{align}
  \langle \bar{\calO}^\alpha_{\Omega_1} \calO^\beta_{\Omega_2} \rangle = &
  \Tr{ \hrho_S \bar{\calO}^\alpha_{\Omega_1} \calO^\beta_{\Omega_2} }
\label{eq:corr_rhoS}
\end{align}
To evaluate this,
let us the reduced density matrix $\hrho_S^{\Omega}$
on the region $\Omega$, where $\Omega = \Omega_1 \cup \Omega_2$, by
\begin{equation}
  \hrho_S^{\Omega} \equiv \Tr_{\bar{\Omega}}{\hrho_S}
\end{equation}
where $\bar{\Omega}$ is the complement of $\Omega$.
We can choose the region $\Omega$ so that $|\Omega|$, in addition to $|\Lambda|$,
is an integral multiple of $|H|$ (and so does $|\bar{\Omega}|$).
Then, using Eq.~\eqref{eq:tru_power},
\begin{align}
  \hrho_S^{\Omega} = & \Tr_{\bar{\Omega}}{\hrho_S}
  \notag \\
  \sim & \frac{1}{d^{|\Lambda|}} \sum_{g \in \tilde{G}} \prod_{i \in \Omega} u_g^{(i)} \prod_{j \in \bar{\Omega}} \tr{u_g^{(j)}} 
  \notag \\
  \sim & \frac{1}{d^{|\Omega|}} \mathbb{I}^{\Omega} ,
\end{align}
Since the reduced density matrix corresponds to a completely disordered state
in which there is no correlation between $\Omega_1$ and $\Omega_2$,
the standard correlation function~\eqref{eq:corr_rhoS} vanishes.


Therefore we need to consider the Rényi-2 correlation function to
characterize $\hrho_S$.
First we discuss the purity.
Again using Eq.~\eqref{eq:tru_power},
\begin{align}
  \Tr{{\hrho_S}^2} \sim &
  \Tr{\frac{1}{d^{2|\Lambda|}} \sum_{g, g' \in \tilde{G}} U_g U_{g'} } 
  \notag \\
 = & \frac{1}{d^{2|\Lambda|}} |\tilde{G}| \sum_{g'' \in G} \Tr{\left[ U_{g''} \right]}
  \notag \\
\sim & \frac{|\tilde{G}|}{d^{2|\Lambda|}} d^{|\Lambda|} = \frac{|\tilde{G}|}{d^{|\Lambda|}} .
\label{eq:purity}
\end{align}
This is $|\tilde{G}|=|G|/|H|$ times larger than the purity of the completely random state.
This can be also seen from Eq.~\eqref{eq:N_id}.

Finally,
the Rényi-2 correlation function is evaluated as
\begin{align}
 & \Tr{\hrho_S \bar{\calO}^\alpha_{\Omega_1} \calO^\beta_{\Omega_2}
     \hrho_S \calO^\alpha_{\Omega_1} \calO^\beta_{\Omega_2} } 
     \notag \\
  & \sim 
  \sum_{g, g' \in \tilde{G}} \frac{1}{d^{2|\Lambda|}}
  \Tr{ U_g \bar{\calO}^\alpha_{\Omega_1} \calO^\beta_{\Omega_2} 
    U_{g'} \calO^\alpha_{\Omega_1} \bar{\calO}^\beta_{\Omega_2} }
  \notag \\
   &\sim
  \frac{1}{d^{2|\Lambda|}} 
  \sum_{g, g' \in \tilde{G}}
\Tr_\Omega{\left[ U^\Omega_g \bar{\calO}^\alpha_{\Omega_1} \calO^\beta_{\Omega_2}
U^\Omega_{g'} \calO^\alpha_{\Omega_1} \calO^\beta_{\Omega_2} \right]} 
    \Tr_{\bar{\Omega}}{U^{\bar{\Omega}}_g U^{\bar{\Omega}}_{g'}} .
\end{align}
 The partial trace $\Tr_{\bar{\Omega}}$ is evaluated as
 \begin{align}
  \Tr_{\bar{\Omega}}{U^{\bar{\Omega}}_g U^{\bar{\Omega}}_{g'}} = & \Tr_{\bar{\Omega}}{U^{\bar{\Omega}}_{gg'}}
\notag \\
\sim d^{|\bar{\Omega}|} \delta_{gg',e} 
\end{align}
for $|\bar{\Omega}|, |\Lambda| \to \infty$ (while keeping $|\Lambda|$ and $|\Omega|$
integral multiples of $|H|$).
Thus only the terms $g'=g^{-1}$ contributes to the sum.
\begin{align}
  &\Tr{\left[ \hrho \bar{\calO}^{\Omega_1} \calO^{\Omega_2} \hrho \calO^{\Omega_1} \calO^{\Omega_2} \right]}   \notag \\
  \sim & \frac{|G|}{d^{|\Lambda|+|\Omega|}} 
  \sum_{g \in \tilde{G}}
     \Tr_\Omega{\left[ {U^\Omega_g}^\dagger \bar{\calO}^{\Omega_1} \calO^{\Omega_2} U_g^\Omega \calO^{\Omega_1} \bar{\calO}^{\Omega_2} \right]} .
\end{align}
This is generally non-vanishing if $\bar{\calO}^{\Omega_1} \calO^{\Omega_2}$ and $\calO^{\Omega_1} \bar{\calO}^{\Omega_2}$ contains the identity representation
of $\tilde{G}$ (or $G$).

\section{Strong-to-weak symmetry breaking criticality}
\label{app:cri}
Generalizing our analysis, we discuss mixed states criticality under decoherence.
We start from the ground state of the Hamiltonian
\begin{align}
H=- \sum_i \left(Z_i \tilde{Z}_{i,i+1} Z_{i+1} +
 \tilde{X}_{i-1,i} X_i  \tilde{X}_{i,i+1}
 + X_i
 \right) ,
\end{align}
which is different from Eq.~\eqref{1Dstabilizer} just by the transverse field term $X_i$.
This is a simple model Hamiltonian for ``gapless SPT'' state~\cite{gaplessSPT}.
In fact, by a duality transformation~\cite{verresen2021efficiently}, it is mapped to
\begin{align}
H=- \sum_i \left(Z_i  Z_{i+1} + X_i \right)
 - \sum_i \left( \tilde{X}_{i-1,i} \tilde{X}_{i,i+1}
 \right) ,
\end{align}
which is a critical Ising chain of $Z_i$ and the classical Ising chain of $\tilde{X}_i$,
and they are decoupled from each other.
The correlation function of the dual model implies the string correlation in the
original model as
\begin{align}
\langle Z_i \left( \prod_{a=0}^{j-i-1} \tilde{Z}_{i+a}  \right) Z_j \rangle 
\sim & \frac{1}{|j-i|^{1/4}} ,
\\
\langle \tilde{X}_i \left( \prod_{a=1}^{j-i} X_{i+a}  \right) \tilde{X}_j \rangle 
\sim &  1 .
\end{align}
The density matrix $\hrho$ for $Z_i$ spins obtained by tracing out ancilla
has no long-range or quasi-long-range order, and the usual correlation functions
vanish.
Nevertheless, the R\'{e}nyi-2 correlator is nonvanishing and inherits
the power-law behavior of the string correlation in the gapless SPT state as
\begin{align}
    \frac{\tr{\left(Z_i Z_j \hrho Z_i Z_j \hrho \right)}}{\tr{\hrho^2}} \sim \frac{1}{|i-j|^{1/2}} .
\end{align}

\end{document}